\documentclass[12pt,a4paper]{article}
\pdfoutput=1
\usepackage{comment}
\usepackage{listings}
\usepackage{amsfonts}
\usepackage{caption}
\usepackage{subcaption}
\usepackage{algorithm}
\usepackage{amsmath} 
\usepackage{graphicx} 
\usepackage{algorithmicx}
\usepackage{algpseudocode}
\usepackage{xcolor}

\begin{document}
\title{Discontinuous Galerkin Methods on Graphics Processing Units for Nonlinear Hyperbolic Conservation Laws}
\author{Martin Fuhry \and Andrew Giuliani \and Lilia Krivodonova}
\date{}
\maketitle

\begin{abstract}
We present a novel implementation of the modal discontinuous Galerkin (DG) method for hyperbolic conservation laws in two dimensions on graphics processing units (GPUs) using NVIDIA's Compute Unified Device Architecture (CUDA).
Both flexible and highly accurate, DG methods accommodate parallel architectures well as their discontinuous nature produces element-local approximations.
High performance scientific computing suits GPUs well, as these powerful, massively parallel, cost-effective devices have recently included support for double-precision floating point numbers.
Computed examples for Euler equations over unstructured triangle meshes demonstrate the effectiveness of our implementation on an NVIDIA GTX 580 device.
Profiling of our method reveals performance comparable to an existing nodal DG-GPU implementation for linear problems.
\end{abstract}

\section{Introduction}

    Graphics processing units (GPUs) are becoming increasingly powerful, with some of the latest consumer offerings boasting over a teraflop of double precision floating point computing power \cite{tesla}.
    Their theoretical computing performance compared with their price exceeds that of nearly any other computer hardware.
    With the introduction of software architectures such as NVIDIA's CUDA \cite{cuda}, GPUs are now regularly being used for general purpose computing.
    With support for double-precision floating point operations, they are now a serious tool for scientific computing.
    Even large-scale computational fluid dynamics (CFD) problems with several million element meshes can be easily handled in a reasonable amount of time as video memory capacities continue to increase.   
    

Solvers operating on structured grids have been implemented with success \cite{brandvik2,stvenant,struc3,michea}.  Compared to structured grid methods, unstructured grid CFD solvers have been sparse \cite{corrigan}.
    It is likely that this is due to several factors.  First, data dependent memory access patterns inherent to unstructured grids generally do not lead to efficient memory bandwidth use.  
    Further, low-order methods commonly do not exhibit sufficient floating point operations to hide long memory access latencies \cite{jacob}.  
    Fortunately, it is possible to overcome these inconveniences by optimizing memory transactions as is done in \cite{corrigan} and \cite{jacob}.  
    High-order methods with a sufficient compute-to-memory access ratio such as Flux Reconstruction (FR) \cite{patrice}, correction procedure via reconstruction (CPR) \cite{cpr,cpr2}, and DG methods  \cite{dggpu1,dggpu2,Siebenborn:2013:GAD:2499968.2499971,sieb2,goedel2,Klöckner2012225,klockner-thesis}
 have also been implemented with success on GPUs.

    
     We describe a GPU implementation of a modal discontinuous Galerkin method for solutions of nonlinear two-dimensional hyperbolic conservation laws on unstructured meshes.
    High-order DG methods can be particularly suitable for parallelization on GPUs as the arithmetic intensity, that is, the ratio of computing time to memory access time, is significantly higher than, e.g., finite difference and finite volume methods due to the higher concentration of degrees-of-freedom (DOF) per element.
    The amount of work per degree-of-freedom is also higher as it requires a computationally intensive high-order quadrature rule evaluation over an element and its boundary.

    A relevant example to this work is Kl\"ockner, Warburton and Hesthaven  \cite{dggpu1,dggpu2}, who implemented a high-order nodal DG-GPU method on a system of linear three dimensional hyperbolic equations.  
    Linear solvers have simplified computations by taking advantage of linearity, i.e., the evaluation of the integrals over cells may be precomputed with the precomputed values reused at each timestep.  
    Characteristic of DG methods, their implementation consists of both a communication step that evaluate fluxes to permit the exchange of information between elements, and element-local steps characterized by matrix-vector multiplications.  
    On the GPU, work was partitioned to thread blocks based on systematic empirical tests, yielding optimal runtimes.
	As a result, the amount of work a block completed was dependent on, for example, the order of approximation of the method and the operation being completed.
	Data was organized such that the degrees-of-freedom for one element were adjacent to one another, i.e. organized element-wise.
    Their code exhibited good utilization of memory bandwidth and impressive arithmetic throughput. 
    They report speed-ups of 40 to 60 relative to an optimized CPU implementation.  In this same vein, a multi-GPU linear nodal DG-GPU implementation for Maxwell's equations can be found in \cite{goedel2}.
   
      Nonlinear solvers have been investigated as well.  
      Siebenborn and Schulz  \cite{Siebenborn:2013:GAD:2499968.2499971}, Siebenborn, Schultz and Schmidt \cite{sieb2} along with Gandham, Medina and Warburton \cite{gandham} use again nodal DG-GPU to solve the Euler and shallow water equations, respectively. 
      Nonlinear solvers, unlike linear ones, must recompute integrals containing nonlinear fluxes for each element at every timestep.  As a result, their implementation requires not only numerical integration, but also interpolation from solution nodes to integration points.  
      Siebenborn et al. and Gandham et al., like Kl\"ockner et al., take the approach of efficiently implementing matrix-vector multiplications, with the addition of parallelized interpolation to surface and volume integration points.  
      It would seem logical to use NVIDIA's cuBLAS for such computations, but it turns out custom implementations are more efficient for both linear and nonlinear solvers \cite{Siebenborn:2013:GAD:2499968.2499971, sieb2,dggpu1,dggpu2,Klöckner2012225}.
      Work partitioning for Siebenborn's implementation \cite{sieb2} assigned one matrix-vector multiplication, i.e. corresponding to one element, to each thread block.  The partitioning of work to individual threads then depended on the dimensions of the matrix involved.
                  A final particularity of nonlinear problems is that shocks and discontinuities may appear in solutions.  Such phenomena can cause spurious oscillations that degrade the quality of the numerical solution.  Adding an artificial viscosity term \cite{persson2006sub} or slope limiting are two possible measures that can be taken.

    An important aspect of both linear and nonlinear DG implementations is the treatment of curvilinear solid boundaries.  
    If not properly accounted for, the accuracy of the numerical solution near such boundaries will be adversely affected; 
    this is particularly true for DG numerical solutions as shown by Bassi and Rebay \cite{bassi1997high}.  
    In order to account for nonlinear boundaries, there are different approaches of varying efficacy and ease of implementation.  
    For the Euler equations, Bassi et al. show that implementing at least a quadratic geometric order of approximation is necessary for curved boundaries.  
    Doing so however requires storing extra information about boundary elements, such as the Jacobian of the nonlinear mapping \cite{hesthaven}.  
    An easier technique proposed by Krivodonova and Berger \cite{boundary} suggests keeping a piecewise linear boundary, but imposing a curvature boundary condition, i.e. at each integration point along a piecewise linear numerical boundary, ensuring streamlines are tangent to the physical boundary.  The latter approach was adopted in this work.
    
    In summary, the main difference between linear and nonlinear DG solvers lies in the ability or inability to precompute the majority of the work.  The unifying aspect of the previous work discussed above is the use of efficient matrix-vector multiplications.  Due to their frequent data reuse, such algorithms use shared memory to reduce accesses to slow global memory.

We diverge from previous implementations in two respects: work partitioning and, consequently, DOF organization.  First, our implementation delegates work in a simple thread-per-element and thread per-edge fashion.  As with previous DG-GPU implementations, our calculations will involve evaluating an integral over an element and evaluating an integral over its boundary.  While the former computation is straightforward to implement, two different approaches may be taken towards implementing the latter.  Element-wise integration over edges results in evaluating the same integral twice.  Edge-wise integration significantly increases thread count and avoids performing this computation twice, but introduces race conditions for unpartitioned, unstructured meshes.  Preventing these race conditions when parallelizing edge-wise proved particularly challenging, as atomic operators significantly degrade performance.  Despite these difficulties, we found that the edge-wise approach provided roughly twice the performance as the element-wise approach.  This approach to work partitioning avoids using shared memory and the access pattern requirements therein.  
Finally, we do not organize solution coefficients, i.e. degrees-of-freedom, element-wise like Kl\"ockner et al. \cite{dggpu1,dggpu2,klockner-thesis}.  Rather, we store these DOFs basis-function-wise to accommodate our particular access patterns.

    In the final section of our paper, we present several representative examples based on solution of the Euler equations with and without limiting.
    We measure runtime and various kernel performance characteristics, along with scalability for order of approximation $p$ ranging from one to five.
    To take full advantage of the processing power available in our GTX 580, we found that a mesh size of roughly 10,000 elements suffices.
    With $p = 1$, we found that problems of up to four million triangles can be easily handled for Euler equations.
    We found that while limiters certainly inhibit performance on GPUs as they do on CPUs, performance in our implementation does not degrade significantly.
    Finally, we conclude that GPU performance comparable to \cite{dggpu1,dggpu2} can be achieved, by operating on a much simpler thread-per-element and thread-per-edge basis without using shared memory between threads.  Our implementation demonstrates that a straightforward, but careful, approach can achieve good device utilization even for complicated problems.

\section{The Discontinuous Galerkin Method}

        The DG method for a two dimensional nonlinear system of $M$ equations is now presented. We are interested in the numerical approximation of
        \begin{align}
            \label{eq:2ddg}
            \partial_t \mathbf{u} + \nabla_{xy} \cdot \mathbf{F}(\mathbf{u}) = \mathbf{0}
        \end{align}
        for a vector $\mathbf{u} = [u_1, u_2, \dots, u_M]$
        over a computational domain $\Omega \subset \mathbb{R}^2$,
        with a sufficiently smooth flux function $\mathbf{F} = [F_1, F_2]$, where $F_1$ and $F_2$ are the fluxes in the $x$ and $y$ directions, respectively.
        We enforce the initial conditions
        \begin{align}
            \mathbf{u}(x, y, 0) = \mathbf{u}_0(x, y),
        \end{align}
        and appropriate boundary conditions.
        We first partition $\Omega$ into a mesh of triangles 
         \begin{align}
            \Omega = \bigcup_{i = 1}^N \Omega_i,
        \end{align}
        then multiply $(\ref{eq:2ddg})$ by a test function $v \in H^1(\Omega_i)$, integrate over the element $\Omega_i$, and use the divergence theorem to obtain the weak formulation
        \begin{align}
            \label{eq:weak2dsys}
            \frac{d}{\,dt} \int_{\Omega_i} v \mathbf{u} \,d\mathbf{x} 
            + \int_{\Omega_i} \nabla_{xy} v \cdot \mathbf{F}(\mathbf{u}) \,d{\mathbf{x}} 
            - \int_{\partial \Omega_i} v \mathbf{F}(\mathbf{u}) \cdot \mathbf{n}_i \,ds = 0,
        \end{align}
        where $\mathbf{n}_i$ is the unit, outward-facing normal for element $\Omega_i$'s edges.
 
        To obtain a simpler formulation, we map each $\Omega_i$ to a canonical triangle $\Omega_0$ with vertices at $(0, 0), (1, 0),$ and $(0, 1)$ with the bijection
        \begin{align}
	\label{eq:2dmap}
            \begin{pmatrix}
                x \\
                y \\
                1
            \end{pmatrix}
            =
            \begin{pmatrix}
                x_1 & x_2 & x_3 \\
                y_1 & y_2 & y_3 \\
                 1   & 1   & 1 
            \end{pmatrix}
            \begin{pmatrix}
                  1 - r - s \\
                  r \\
                  s
            \end{pmatrix},
        \end{align}
        where $(x_k, y_k),~ k = 1, 2, 3$, are the vertices of the given element and $\mathbf{r} = (r,s) \in \Omega_0$.
        We arrange the ordering of the vertices of $\Omega_i$ in a counter-clockwise direction to enforce a positive determinant of the Jacobian  $J_i$ of the transformation (\ref{eq:2dmap}).
        With this mapping, the integral of the flux over $\Omega_i$ in the weak formulation (\ref{eq:weak2dsys}) over element $\Omega_i$ becomes
        \begin{align}
            \int_{\Omega_i} \nabla_{xy} v \cdot \mathbf{F}(\mathbf{u}) \,d{\mathbf{x}} = \int_{\Omega_0} (J^{-1}_i \nabla v) \cdot \mathbf{F}(\mathbf{u})\det{J_i}\,d\mathbf{r},
        \end{align}
        which we refer to as the volume integral.  The gradient with respect to the reference coordinate system $(r,s)$ is denoted simply by $\nabla = (~\frac{\partial}{\partial r}~,~ \frac{\partial}{\partial s}~)$.
        
        We separately map each edge of $\Omega_i$ to the canonical interval $I_0 = [-1 ,1]$ by the bijective mapping given by
        \begin{align}
            \label{eq:1dmap}
            \begin{pmatrix}
                x \\
                y 
            \end{pmatrix}
            =
            \begin{pmatrix}
                x_1 & x_2 \\
                y_1 & y_2
            \end{pmatrix}
            \begin{pmatrix}
                \frac{1}{2}(1 - \xi) \\
                \frac{1}{2}(1 + \xi) 
            \end{pmatrix},
        \end{align}
        where $(x_k, y_k),~k=1,2,$ are the endpoints of the given edge and $\xi \in I_0$.
           Using this mapping to $I_0$, the integral over $\partial \Omega_i$ in the weak formulation (\ref{eq:weak2dsys}) becomes
            \begin{align}
                \label{eq:surfaceintegral}
                \int_{\partial \Omega_i} v \mathbf{F}(\mathbf{u}) \cdot \mathbf{n}_i \,ds = 
                        \sum_{q = 1}^3 \int_{I_0} v_q \mathbf{F}(\mathbf{u}) \cdot \mathbf{n}_{i,q} l_{i,q} \,d\xi,
            \end{align}
        where $\mathbf{n}_{i,q}$ denotes the unit outward-facing normal vector for edge $q$, $v_q$ is the test function $v$ restricted to that edge, and $l_{i,q}$ denotes the determinant of the Jacobian of (\ref{eq:1dmap}), given by
        \begin{align}
            \label{eq:len}
            l_{i,q} = \frac{1}{2}\sqrt{(x_1 - x_2)^2 + (y_1 - y_2)^2}.
        \end{align}
        We refer to (\ref{eq:surfaceintegral}) as the surface integral.  

        \begin{figure}[tbp]
            \caption{Basis function $\phi_5$ evaluated on $\Omega_0$ (left) and $\partial \Omega_0$ (right)}
            \label{fig:basisedge}
            \centering
            \begin{subfigure}[b]{.4\textwidth}
                \includegraphics[width=\textwidth]{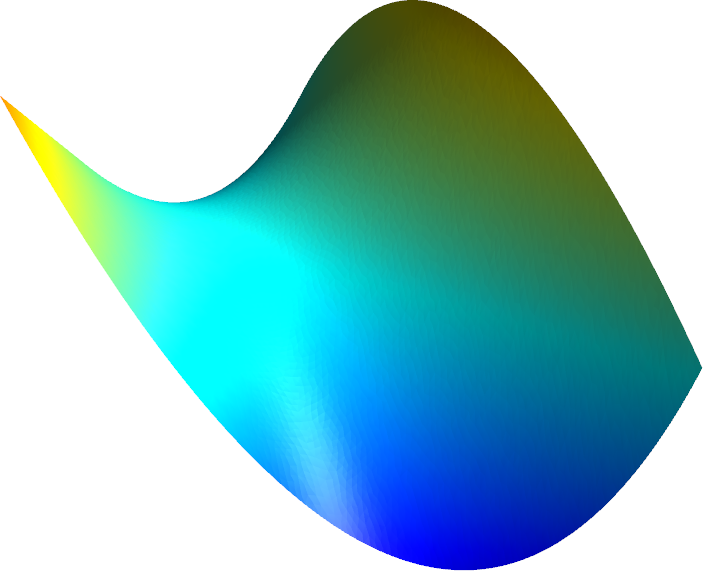}
                \caption{$\phi_5$ on $\Omega_0$}
            \end{subfigure}
            ~
            \begin{subfigure}[b]{.5\textwidth}
                \includegraphics[width=\textwidth]{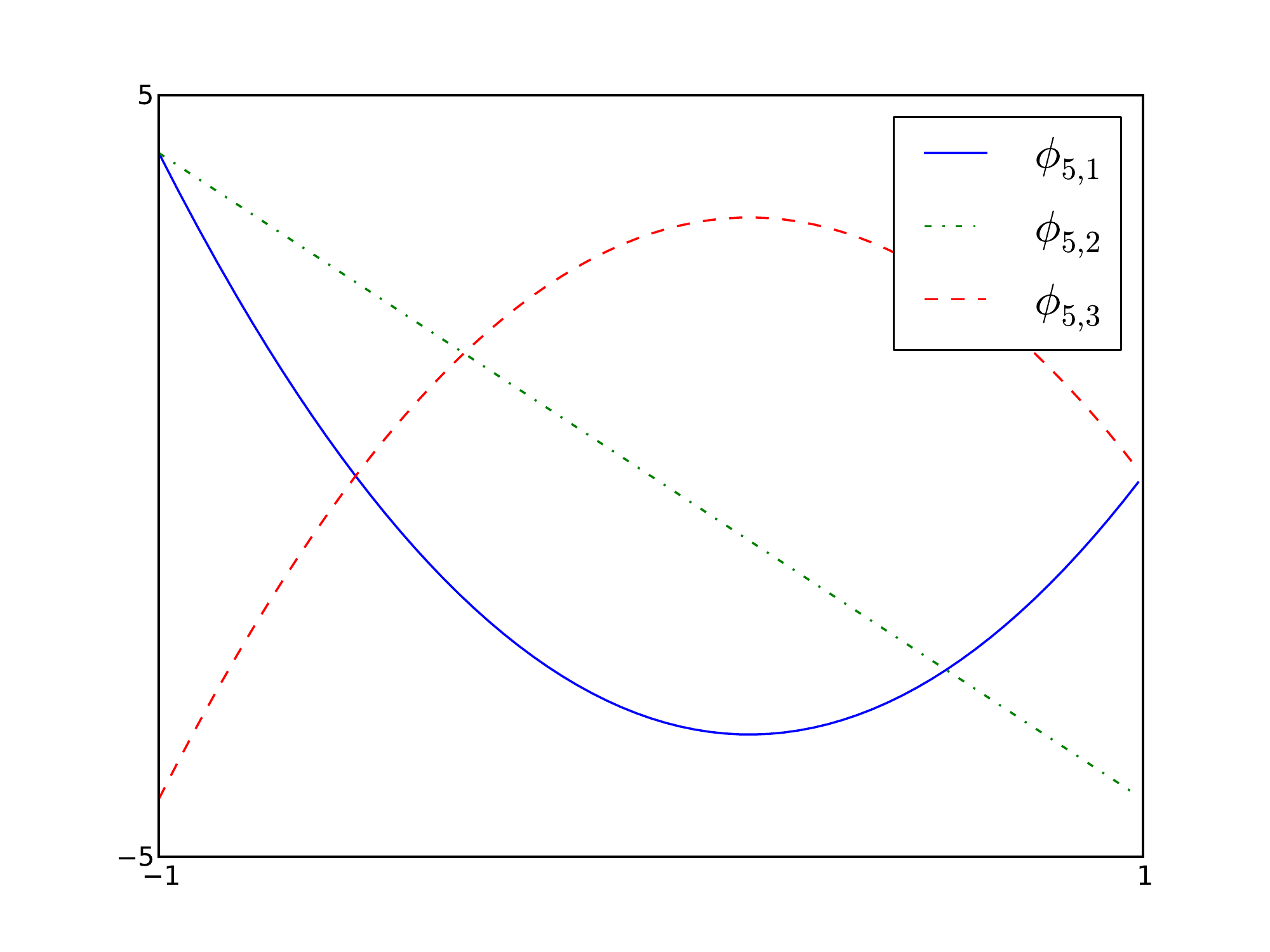}
                \caption{$\phi_5$ evaluated over $\partial \Omega_0$}\label{fig:basisedgeb}
            \end{subfigure}
        \end{figure}

        Let $\Phi = \left\{\phi_j\right\}_{j=1}^{N_p}$ be an orthonormal basis for the space of polynomials of degree at most $p$ on $\Omega_0$, $S^p(\Omega_0)$, consisting of
        \begin{align}
            \label{eq:np}
            N_p = \frac{1}{2}(p + 1)(p + 2)
        \end{align}
        basis functions given in, e.g., \cite{polynomials}.
        Let $\mathbf{U}_i$ approximate $\mathbf{u}$ over $\Omega_i$ using a linear combination of basis functions $\phi_j$ ,
        \begin{align}
            \mathbf{U}_i = \sum_{j=1}^{N_p} \mathbf{c}_{i,j}(t)\phi_j,
        \end{align}
        where $\mathbf{c}_{i,j}(t) = [c^1_{i,j},c^2_{i,j},\cdots, c^M_{i,j}]$ is a vector of solution coefficients.
        Choosing each test function $v = \phi_j \in \Phi$ $j = 1,2,\hdots,N_p$ in (\ref{eq:weak2dsys}) and using the orthonormality of the basis creates the system of ODEs
        \begin{align}
            \label{eq:2ddg0}
            \frac{\,d}{\,dt} \mathbf{c}_{i,j}(t) = \frac{1}{\det{J_i}}\left(\int_{\Omega_0} \mathbf{F}(\mathbf{U}_i) \cdot (J^{-1}_i \nabla \phi_j) \,\det{J_i}d\mathbf{r} 
                     - \sum_{q = 1}^3 \int_{I_0} \phi_{j,q} \mathbf{F}(\mathbf{U}_i) \cdot \mathbf{n}_{i,q} l_{i,q} \,d\xi\right),
        \end{align}
        where $\phi_{j,q}$ denotes the basis function $\phi_j$ evaluated on the edge $q$, as demonstrated in Figure \ref{fig:basisedge}.
        Note that in general $\phi_j$ assumes different values on each edge $q$ (Figure \ref{fig:basisedgeb}).  
        Because of the orthonormality of the basis functions, the mass matrix is the identity matrix.       
        The numerical solution on the edges of each element is twice defined, as we do not impose continuity between elements.
        To resolve these ambiguities, we replace the flux $\mathbf{F}(\mathbf{U}_i)$ on $\partial \Omega_i$ with a numerical flux function $\mathbf{F_n}(\mathbf{U}_i, \mathbf{U}_k)$, using information from both the solution $\mathbf{U}_i$ on $\Omega_i$ and the solution $\mathbf{U}_k$ of the neighboring element $\Omega_k$ sharing that edge.
        Finally, (\ref{eq:2ddg0}) can be integrated in time using an ODE solver, e.g. a Runge-Kutta method. 
        
\section{CUDA Programming Paradigm}
        GPU computing, while providing perhaps the most cost-effective performance to date, is not without its challenges.
        The sheer floating point power dispensed by GPUs is only realized when unrestricted by the GPUs limited memory bandwidth.
        As such, it is of paramount importance to carefully manage memory accesses. There are many different types of memory available to the GPU programmer.  Of interest to us are global, thread private, constant, and shared memories, the first three of which are used in our implementation.

        Global memory on the GPU is very large and located in GPU video (DRAM) memory.  Though this memory is cached, on a cache miss, it is very slow and introduces large latencies.
        We therefore manage global memory by minimizing these memory accesses, coalescing global memory transactions by instructing nearby
threads to access the same blocks of memory.
        Thread private memory can be located in either fast registers or, in the case of register spilling, in global memory.  In the latter case, data is cached, permitting faster access.
        Constant memory is cached and extremely fast, but very limited.
        Finally, shared memory is convenient when there is frequent data reuse, as in the case of multiple matrix-vector multiplications.  This type of memory is shared amongst a group, or block, of threads and is faster than accessing data from global memory, provided bank conflicts are avoided \cite{cuda}.
        Although we do not use shared memory as do many nodal DG-GPU implementations \cite{dggpu1,dggpu2,Siebenborn:2013:GAD:2499968.2499971,sieb2, klockner-thesis,gandham}, precomputed data subject to frequent reuse will be stored in constant memory.

        Programming GPUs using CUDA is unlike programming CPUs \cite{cuda}.  In this programming model, the CPU is known as the host, whereas the GPU is known as the device.
        CUDA machine code is separated into kernels - parallel algorithms to be executed on the GPU in SIMD (single instruction, multiple data) fashion or in lock-step.  Our implementation uses both \lstinline|global| kernels, launched solely by the host, and \lstinline|device| kernels launched solely by the device, i.e. by active global kernels.
        
        Programming is done from a thread perspective, as each thread reads the same instruction set, using it to manipulate different data.
        Threads have a unique numeric identifier, a thread index, allowing, e.g., each thread to read from a different memory location.
        In order to take full advantage of the processing power available, enough warps, that is, collections of thirty-two threads, must be created to fully saturate the device.
        Load balancing is done entirely by CUDA's warp scheduler.
        When memory access requests introduce latency, e.g., global memory access requests, the warp scheduler will attempt to hide the latency by swapping out the idle warps and replacing them with ready warps.
        Problems with low arithmetic intensity spend most of their runtime waiting during latencies in memory transactions.
        Problems with high arithmetic intensity, on the other hand, allow CUDA's warp scheduler to hide memory latency behind computation by scheduling ready warps ahead of those waiting on memory transactions.

        Boolean expressions pose problems for GPUs, as two threads in the same warp may evaluate a boolean condition to different values.
        When this happens, the warp splits into branches, where each branch of threads executes a different path of code split by the boolean.
        This branching, called warp divergence, harms performance as both instructions must be run for this warp.

        Below, we give an overview of our parallelization approach for this implementation within the constraints of the CUDA programming model. 
        The following subsections provide more detailed descriptions of our method.
            
\section{Implementation}

        We now outline the steps taken to parallelize the evaluation of the right-hand side of equation (\ref{eq:2ddg0}).  This computation combines a volume integral
        \begin{align}
            \label{eq:volume}
            \int_{\Omega_0} \mathbf{F}(\mathbf{U}_i) \cdot (J^{-1}_i \nabla \phi_j) \,\det{J_i}d\mathbf{r} 
        \end{align}
        and a surface integral
        \begin{align}
            \label{eq:surface}
            \sum_{q = 1}^3 \int_{I_0} \phi_{j,q} \mathbf{F}(\mathbf{U}_i) \cdot \mathbf{n}_{i,q} l_{i,q} \,d\xi
        \end{align}
        consisting of three independent line integrals, for each $j = 1,2, \dots, N_p$. 
        Parallelization is therefore straightforward.
        As the volume integral contributions (\ref{eq:volume}) for one element require only local information from that element, i.e., the element's coefficients, the determinant of the Jacobian and its inverse, these contributions on each element can be computed independently.
        Similarly, each edge's surface integral contributions (\ref{eq:surface}) are computed independently as these computations require local information from two elements sharing that edge, i.e., both elements' coefficients.
        
        An explicit integration scheme advances the solution coefficients in time.

    \subsection{Data Structure and Memory Management} \label{sec:ds}

In this section we describe the major data structures employed in this implementation.  We begin with the structure of the solution coefficient array, followed by precomputed data and mesh connectivity.  First, the solution coefficients of equation (\ref{eq:2ddg0}) are stored in an array $\mathbf{C}$.  This array is located in global memory and organized as follows

        \begin{figure}[H]
            \caption{Organization of solution coefficients in $\mathbf{C}$}
            \label{fig:coalesce}
                \centering
                \includegraphics{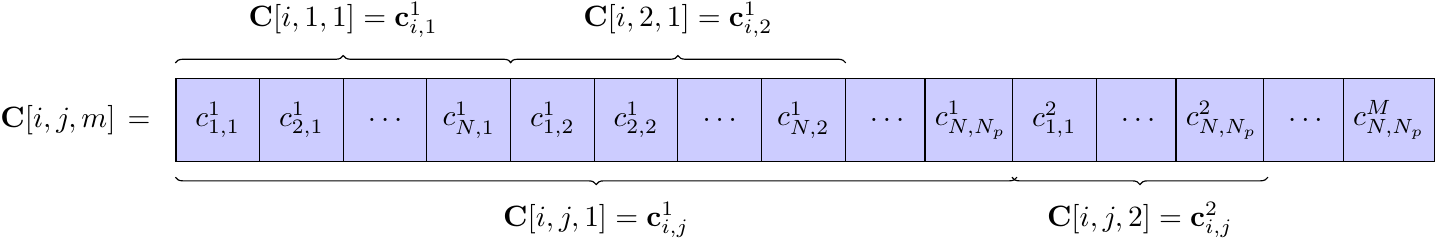}
        \end{figure}

\noindent where $c^m_{i,j}$ is the coefficient corresponding to the $j$th basis function $\phi_j$, for element $i$, of equation $m$.
This organization places the coefficients for one order, i.e. corresponding to one basis function, side-by-side.  
Coalesced memory accesses of $\mathbf{C}$ were attempted as much as possible in order to maximize useful memory throughput.  In order to assure coalesced transactions, nearby threads in a warp must access nearby addresses in memory, i.e. thread $i$ accesses memory address $k$, thread $i+1$ accesses memory address $k+1$.  
In Figure \ref{fig:coalesce}, the predominant access pattern of  $\mathbf{C}$ is shown; each thread $i$ accesses the solution coefficients for an element $\Omega_i$.  They complete a nested for-loop whereby the inner loop first iterates through the basis functions $j = 1, \hdots, N_p$, and the outer loop iterates through equations $m = 1, 2, \hdots, M$.  Unlike other DG-GPU implementations \cite{dggpu2,dggpu1,Siebenborn:2013:GAD:2499968.2499971,sieb2}, we do not pad variables to ensure aligned memory accesses; for Fermi architecture GPUs, the caching of global memory largely mitigates the nefarious effects of misaligned reads \cite{cudafortran}.   Further, the slight performance gain of aligned reads is not judged worthwhile over the possibility of wasted memory due to padding.

        \begin{figure}[H]
            \caption{Coalesced memory access pattern}
            \label{fig:coalesce}
                \centering
                \includegraphics{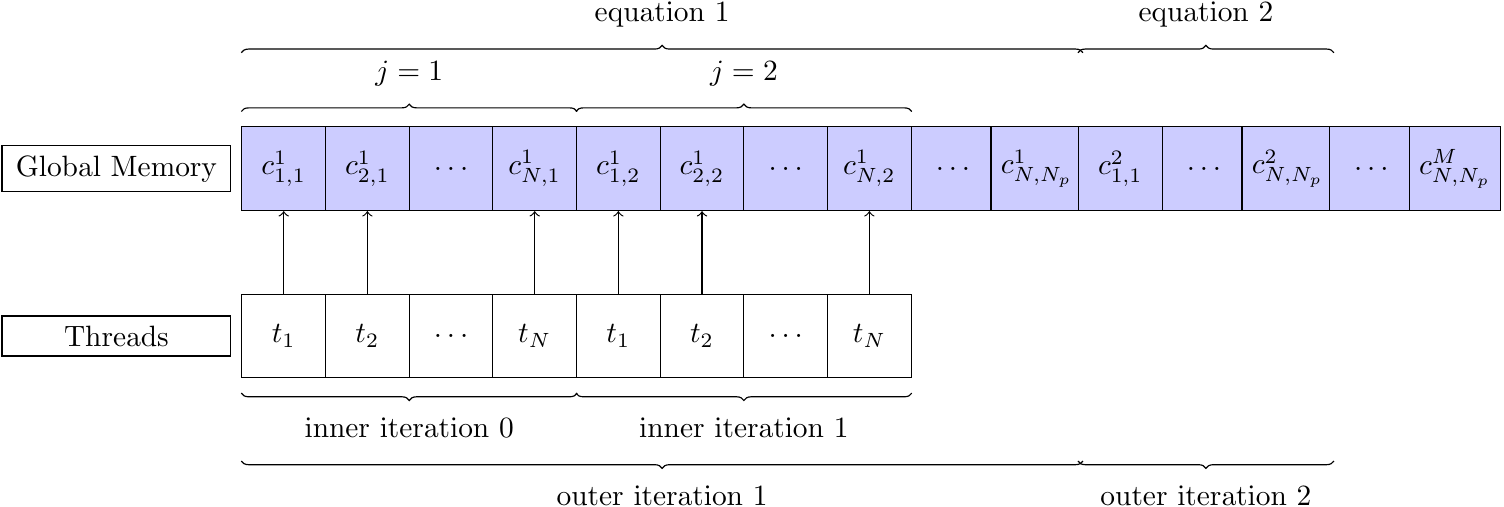}
        \end{figure}

                Other data in global memory is also stored in a manner enabling coalesced accesses. The precomputed matrix $J_i^\tau = |J_i| J_i^{-1}$ and mesh connectivity information for elements are sorted element-wise.  Edge normals, the determinants of the edge mappings' Jacobians (\ref{eq:1dmap}), and mesh connectivity information for edges are sorted edge-wise.
                
\begin{figure}[tpb]
            \caption{Storing the integration points on $\Omega_0$}
            \begin{subfigure}[h!]{0.4\textwidth}
                \caption{The integration points for the interior of $\Omega_0$, $\mathbf{r}_k$, stored in an array}
                \label{fig:interiorquad}
                \centering
                \includegraphics[width=\textwidth]{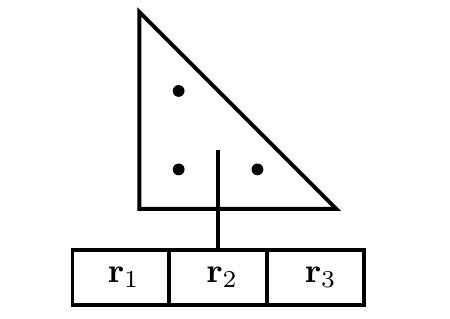}
            \end{subfigure}
            ~
            \begin{subfigure}[h!]{0.6\textwidth}
                \caption{The boundary integration points for $\partial \Omega_0$, $\mathbf{r}_{q,k}$ stored in an array}
                \label{fig:boundaryquad}
                \centering
                \includegraphics[width=\textwidth]{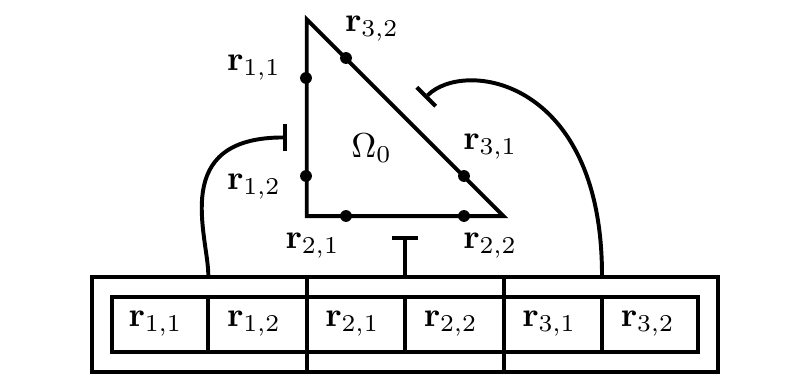}
            \end{subfigure}
        \end{figure}

Constant memory accesses are optimal when every thread in a warp accesses the same address.  
If this is the case, only one read from the constant cache or, on a cache miss, one read from global memory is needed.  
In the worst case, every thread accesses a different memory address, and constant memory reads are serialized \cite{bestpractices}.  
In our implementation, all threads in a warp frequently access the same quadrature rule or basis function value; 
it is clear therefore that read-only constant memory is appropriate for these data.  Thus, we precompute values of $\phi_j(\mathbf{r}_k)$ and $\nabla \phi_j(\mathbf{r}_k)$ at each integration point $\mathbf{r}_k$ in the interior of $\Omega_0$ shown in Figure \ref{fig:interiorquad}, and store them in constant memory.
        We do the same for the values of $\phi_j(\mathbf{r}_{q,k})$ at the integration points $\mathbf{r}_{q,k}$ on $\partial \Omega_0$, shown in Figure \ref{fig:boundaryquad}, where $q = 1,2,3$ denotes which side of the canonical triangle that the integration points reside on.
        These variables are stored as linear arrays: $\mathbf{\Phi_k}$, $\mathbf{\nabla\Phi_k}$, and $\mathbf{\Phi_{q,k}}$ respectively.  Note for conciseness the gradients of the basis functions will be collectively referenced with $\mathbf{\nabla\Phi_k}$, though it is understood that the partial derivatives of the basis functions with respect to $r$ and $s$ at each integration point are stored in separate linear arrays.

GPU constant memory is easily able to store the quantities mentioned above for practical values of degree of approximation $p$.
        For example, with $p = 5$, we store 2,268 doubles for the precomputed data, which occupies only twenty-nine percent of the available constant memory space on NVIDIA Fermi architectures.

                     The total memory required for computation depends on four factors.
             First, the size of the mesh determines the number of elements ($N$) and edges ($N_s$).  This affects the number of element vertices in global memory.
             Second, the degree of the polynomial approximation $N_p$ determines the number of coefficients required to approximate the solution.
             Third, the size of the system, $M$, requires a vector of solution coefficients for each variable in that system.
             For each element, we require $M \times N_p$ coefficients to represent the approximated solution over that element.
             Additionally, the ODE solver typically needs extra storage variables for intermediate steps or stages, which must be stored in global memory.   
             
             Finally, Fermi architecture GPUs have 64 kB of memory to be used for the L1 cache and shared memory per streaming multiprocessor \cite{cudafortran}.  As we do not use shared memory and use local memory extensively, kernels are initialized with the \lstinline|cudaFuncCachePreferL1| cache configuration.  This yields 48 kB of L1 cache and 16 kB of unused shared memory.

 \subsection{Mesh Connectivity}
\begin{figure}[tpb]
            \centering
            \caption{The mapping between cells and edges for a simple mesh.}
            \begin{subfigure}[h!]{0.4\textwidth}
                \centering
                \includegraphics[width=\textwidth]{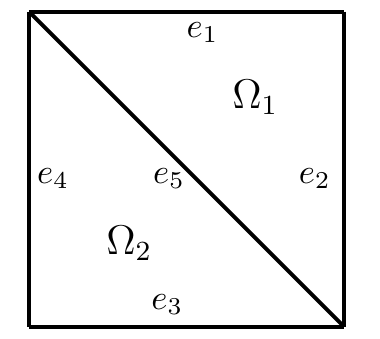}
            \end{subfigure} 
            ~
            \begin{subfigure}[h!]{0.4\textwidth}
                \centering
                \includegraphics[width=\textwidth]{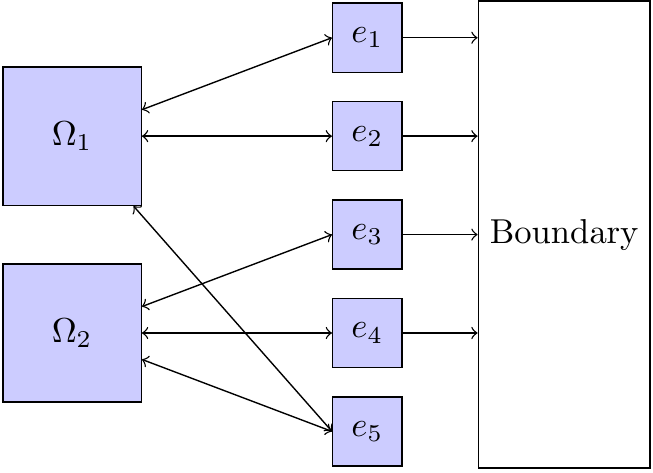}
            \end{subfigure}
            \label{fig:datamappings}
        \end{figure}

        Figure $\ref{fig:datamappings}$ shows an example of a simple mesh consisting of two elements.
        Elements $\Omega_1$ and $\Omega_2$ point to their respective edges; edges $e_1, \dots, e_5$ point back to their respective elements. 
        Each edge stores two pointers: one each to its left and right elements, $\Omega_l$ and $\Omega_r$.
        We arbitrarily assign each of the two elements sharing an edge as either left or right.
        Edges lying on the boundary of the domain have only a left element and store a negative integer describing the type of boundary conditions assigned to them in place of a pointer to a right element.
        The normal vector belonging to an edge points from $\Omega_l$ to $\Omega_r$, by our convention.
        For edges lying on the boundary of the domain, the normal vector therefore points outward.

    \subsection{Outline of Algorithm}\label{sec:volume}
    In this section we outline the algorithm devised to solve (\ref{eq:2ddg0}).  We include a brief textual description with our implementation considerations and a detailed pseudocode of the most important kernels.  In doing the latter, we attempted to be as concise as possible, all the while giving an accurate representation of the organization and control structures of our implementation.
    
    Certain notation was adopted for clarity.  A variable's location in memory is described using the conventions outlined in Table \ref{table:convention}, global and device kernels are distinguished with superscripts $k$ and $d$, respectively, and  $[a..b]$ is the set of integers $\{a, a+1, ..., b\}$.
    
    The access of a specific value in an array is given by indices in square brackets, e.g. ${}^g\mathbf{C}[i,j,m]$ accesses coefficient $c^m_{i,j}$, the $j$th solution coefficient of element $i$ of equation $m$.  Assignments with subscript indices are evaluated for all elements of that subscript, e.g. ${}^{t}\mathbf{u_r}[m]\underset{ (m)}\gets {}^d\text{\lstinline|eval_boundary|}(\mathbf{{}^{c}r_{q_r,p+1-k}})$  returns for all equations $m$.  Assignments between global memory and thread private memory, i.e. ${}^t\mathbf{c}[j,m]\underset{(j,m)}\gets{}^g\mathbf{C}[i,j,m]$ are implemented as coalesced accesses as described in Section \ref{sec:ds} where possible.  
     Vector-vector dot product with a subscript like ${}^{t}\mathbf{c_i}[j,m]\underset{ (j)}\cdot {}^{c}\mathbf{\Phi_{k}}[j]$ are completed over all the elements of $j$ for a particular $m$, i.e. ${}^{t}\mathbf{c_i}[1,m]{}^{c}\mathbf{\Phi_{k}}[1] +  {}^{t}\mathbf{c_i}[2,m]{}^{c}\mathbf{\Phi_{k}}[2] +\hdots + {}^{t}\mathbf{c_i}[N_p,m]{}^{c}\mathbf{\Phi_{k}}[N_p]$, and are implemented with for-loops.
    
        \begin{table}[tpb]
            \caption{Variable and function qualifiers}
            \label{table:convention}
            \centering
            \begin{tabular}{|c|c|}
                \hline
                 Qualifier & Meaning\\
                \hline 
                ${}^g\mathbf{C}$ & superscript $g$ - variable in global memory \\ 
                ${}^t\mathbf{c}$ & superscript $t$ - variable in thread private memory \\ 
                ${}^c\mathbf{\Phi}$ & superscript $c$ - variable in constant memory \\ 
                ${}^k\text{\lstinline|eval_volume|}$ & superscript $k$ - global kernel \\ 
				${}^d\text{\lstinline|eval_boundary|}$ & superscript $d$ - device kernel \\ 
                
                \hline
            \end{tabular}
            \end{table}

    \subsubsection{Time-stepping}

In order to integrate the semi-discrete equation (\ref{eq:2ddg0}) in time, we use classic explicit Runge-Kutta (RK) time integrators of order two and four.  This algorithm is directed by the host and requires the evaluation of various stages of the RK timestepper.
In order to do this, we employ three main compute kernels, ${}^k$\lstinline|eval_volume|, ${}^k$\lstinline|eval_surface| and ${}^k$\lstinline|eval_rhs|.  
The first two return the volume and surface integrals in (\ref{eq:volume}) and (\ref{eq:surface}), respectively.  
The third then appropriately sums all surface and volume contributions, returning the right-hand side of (\ref{eq:2ddg0}) to be used by the time-stepper.

As we are implementing a nonlinear solver, we limit both the intermediate RK stages and the solution at the next time level using the kernel ${}^k$\lstinline|limit_c|.  In the coded implementation of the algorithm, we parallelize both vector sums and the calculation of a stable timestep.  Finally, we explicitly unroll the for-loop calculating the intermediate RK stages to reduce unnecessary instructions being executed at each timestep.

We now present the details of the four main kernels mentioned above.    Note that the number of threads per block was chosen to yield the highest speed, not necessarily the highest occupancy.

    \subsubsection{Volume Integration Kernel (\lstinline|eval_volume|)}\label{sec:volume}
        Parallelizing the volume integral contribution (\ref{eq:volume}) computation is straightforward.
        The integration over $\Omega_i$ requires only local information, i.e., the coefficients ${}^g\mathbf{C}[i,j,m]$ where $j=[1..N_p], ~m = [1..M]$, and the precomputed matrix ${}^gJ_i^\tau = (\det{J_i}) J_i^{-1}$ for that element.  We thus create one thread for each $\Omega_i$, $i\in[1,N]$, tasked with computing (\ref{eq:volume}) over that element.
        
        Algorithm \ref{alg:eval_volume} presents the volume integral computations in pseudocode.
        Thread $t_i$ loads the required coefficients into ${}^t\mathbf{c_i}[j,m]$, $j = [1..N_p],~m = [1..M]$, and the matrix into ${}^tJ_i^{\tau}$.
        It evaluates $\mathbf{U}_i$ and the flux $\mathbf{F}(\mathbf{U}_i)$ at the interior integration points $\mathbf{r}_k$.
        The result is then multiplied by each ${}^tJ_i^\tau{}^c\nabla \mathbf{\Phi_k}[j],~j=[1..N_p]$, which is used to compute the numerical integral over the element.  Finally, this is added to a right-hand side storage variable ${}^g$\lstinline|rhs_volume|.
        
        Approximation of the integral (\ref{eq:volume}) is done using numerical integration rules in \cite{gaussian} of order $2p$.
        
        For all orders of approximation $p$, ${}^k$\lstinline|eval_volume| uses 43 registers and an amount of local memory dependent on $p$; see Section \ref{sec:occupancy}, Table \ref{table:occupancy}.

       \begin{algorithm}[H] 
            \caption{\lstinline|eval_volume|}
            \label{alg:eval_volume}
            \begin{algorithmic}[1]
            \Require A grid of $\lceil N/256 \rceil$ blocks, with $256$ threads per block.  Each thread $i$ computes on one element $\Omega_i$.
            \Require Global Memory Inputs: thread $t_i$ receives the solution coefficients on $\Omega_i$ ${}^g\mathbf{C}[i,j,m]$ and ${}^gJ_i^{\tau}$
            \Require Constant Memory Inputs: thread $t_i$ receives ${}^c\mathbf{\Phi_k}$, $ {}^c\mathbf{\nabla\Phi_k}$, and volume quadrature weights ${}^cw_k$.
            \Ensure Outputs: the volume integral contribution for element $\Omega_i$ 
\State {$i \gets $ thread index}
            \State ${}^{g}\text{\lstinline|rhs_volume[i,j,m]|} \underset{ (j,m)}\gets 0$
            \State ${}^{t}\mathbf{c_i}[j,m]\underset{ (j,m)}\gets {}^{g}\mathbf{C}[i,j,m]$

            \\
			\State Load ${}^gJ_i^{\tau}$ into thread private memory
            
            \ForAll { volume integration points $k$ of $\Omega_i$}
                        \State ${}^{t}\mathbf{u_i}[m]\underset{ (m)}\gets {}^{t}\mathbf{c_i}[j,m]\underset{ (j)}\cdot {}^{c}\mathbf{\Phi_{k}}[j]$
              
 \State ${}^{t}\mathbf{F}[m] \underset{(m)} \gets {}^d\text{\lstinline|flux|}({}^{t}\mathbf{u_i})$
 
            \State ${}^{g}\text{\lstinline|rhs_volume[i,j,m]|} \underset{ (j,m)}\gets {}^{g}\text{\lstinline|rhs_volume[i,j,m]|} + {}^cw_k{}^{t}\mathbf{F}[m]\cdot({}^tJ_i^{\tau} {}^c \mathbf{\nabla\Phi_k}[j])$
          \EndFor

            \end{algorithmic}
            
        \end{algorithm}

    \subsubsection{Surface Integration Kernel (\lstinline|eval_surface|)}\label{sec:surface}
        Parallelizing the surface integral contribution  (\ref{eq:surface}) computation proves more challenging.
        There are two approaches to evaluating the surface integral contribution in (\ref{eq:2ddg0}): element-wise and edge-wise.
	  In the element-wise approach, one thread is created for each element to evaluate (\ref{eq:surface}).
	  Alternatively, in the edge-wise approach, one thread evaluates the surface integral over a single edge, adding the resulting contributions to the elements sharing that edge.
	  As expected, the latter approach was found to be about twice as fast as an extra unnecessary flux evaluation is avoided.
         We thus create one thread $t_i$ for each edge $e_i$, shared by elements $\Omega_l$ and $\Omega_r$ to compute 
		\begin{align}
			\label{eq:edgeint}
			\int_{I_0} \phi_{j,q} \mathbf{F}(\mathbf{U}_l, \mathbf{U}_r) \mathbf{n}_i l_i \,ds = \begin{cases} \int_{I_0} \phi_{j,L} \mathbf{F}(\mathbf{U}_l, \mathbf{U}_r) \mathbf{n}_i l_i \,ds  \text{ on }\Omega_l,~L\in\{1,2,3\} \\ \int_{I_0} \phi_{j,R} \mathbf{F}(\mathbf{U}_l, \mathbf{U}_r) (-\mathbf{n}_i) l_i \,ds \text{ on }\Omega_r,~R\in\{1,2,3\}
			\end{cases}
		\end{align}

		The integral (\ref{eq:edgeint}) contributes to the evolution in time of the coefficients $\mathbf{c}_{l,j}$ and $\mathbf{c}_{r,j}$, see equation (\ref{eq:2ddg0}).  However, its contribution is not necessarily the same for $\Omega_l$ and $\Omega_r$.
		This is because  $\phi_{j,q}$ may take different values at the left and right of the same integration point.
		As our implementation supports any unstructured triangular mesh, elements $\Omega_l$ and $\Omega_r$ may map the same edge to different sides of the canonical triangle. 
        For example, $\Omega_l$ may map edge $e_i$ to the side defined by (0, 1), (0, 0), i.e. $q = 1$, while $\Omega_r$ maps that same edge to the side defined by (1, 0), (0, 1), i.e. $q=3$ (Figure \ref{fig:boundaryquad}).
        In this case, the values of the basis function evaluated on side 1 is not equal to the values of the same basis function evaluated on side 2; refer to Figure \ref{fig:basisedge}.
        Thus, we need to know which side of the canonical triangle $e_i$ becomes for each $\Omega_l$ and $\Omega_r$.  This information is precomputed and stored in two identifiers taking the values $1,2$ or $3$ for the first, second and third canonical side, respectively.
        These identifiers, which we call side mappings, are denoted in (\ref{eq:edgeint}) by subscripts $L$ and $R$.
        Note that the rest of the components of the integrand in (\ref{eq:edgeint}), i.e. the numerical flux $\mathbf{F}(\mathbf{U}_l, \mathbf{U}_r)$, edge normal $\mathbf{n}_i$ and determinant of the Jacobian of the edge mapping $l_i$ can be computed only once and therefore reused.  
        This is computationally efficient especially since evaluation of $\mathbf{F}(\mathbf{U}_l, \mathbf{U}_r)$ may represent the most expensive computation in (\ref{eq:edgeint}).   We precompute and store the values of $\mathbf{n}_i$ and $l_i$ in GPU global memory, sorted edge-wise for coalesced access.

                 The surface integral is approximated using Gauss-Legendre quadrature rules of order $2p + 1$ that require $p+1$ integration points and weights.
        We precompute three matrices $(\mathbf{\Phi_q})_{j,k} = \phi_j(\mathbf{r}_{q,k})$ for each side $q = 1,2,3$, of the canonical triangle and store them row-by-row in a single flattened array in GPU constant memory.
        By using each edge's two side mapping indices ($L$ and $R$) as an offset, we are able to lookup the correct integration points to use while avoiding boolean evaluations, which would create warp divergence.
        
\begin{figure}[tbp]
            \caption{The integration points for $\Omega_l$ and $\Omega_r$ must be traversed in opposite directions i.e. the counterclockwise direction for $\Omega_l$ (left plot) and in the clockwise direction for $\Omega_r$ (right plot)}
            \label{fig:twoboundaryquad}
            \centering
            \includegraphics[width=\textwidth]{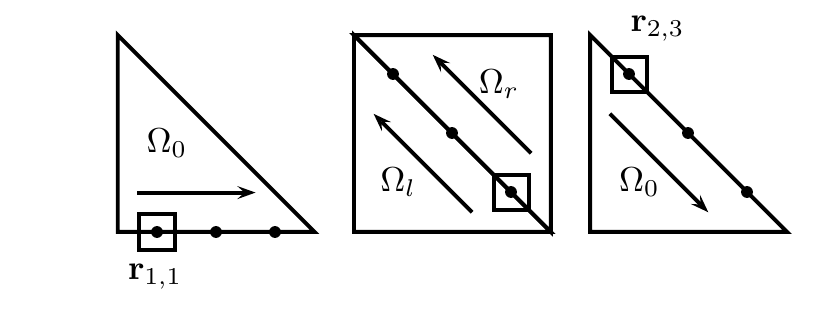}
        \end{figure}

         One final detail we emphasize involves the ordering of our boundary integration points.
         As discussed in Section \ref{sec:ds}, we map integration points from $I_0$ to their respective sides of the canonical triangle in a counter-clockwise direction (Figure \ref{fig:boundaryquad}).
        In Figure \ref{fig:twoboundaryquad}, we illustrate a simple mesh of two elements with three Gauss integration points, indicated with dots, located on the shared edge (see the middle plot).  The mappings of $\Omega_l$ and $\Omega_r$ to $\Omega_0$ are shown in the left and right plots, respectively.  We observe that the shared edge is mapped to side 2 and side 3 of $\Omega_0$ for $\Omega_l$ and $\Omega_r$, respectively.
        Recall that by our convention, the orientation of vertices and integration points on each physical element is counterclockwise.
        Thus, the surface integration points are traversed in opposite directions when viewed from these two elements, i.e. in the counterclockwise direction for $\Omega_l$ and in the clockwise direction for $\Omega_r$.
        
                In Figure \ref{fig:twoboundaryquad}, we enclose with a square the same integration point on the physical edge and on the two canonical edges.
        We see that the first integration point for $\Omega_l$ corresponds to the last integration point for $\Omega_r$, i.e. in order to access the same physical point, we must reverse the direction that we traverse the integration points on $\Omega_r$.  Thus, the flux evaluation at the $k$'th integration point is 
        \begin{align}
            \mathbf{F_n}(\mathbf{U}_l(\mathbf{r}_{L,k}), \mathbf{U}_r(\mathbf{r}_{R,p + 1 - k})).
        \end{align}

        Race conditions prevent us from simply adding the resulting surface integral contributions together with the volume integrals for $\Omega_l$ and $\Omega_r$ as we compute them. 
        For example, two threads assigned to two edges belonging to the same $\Omega_i$ may compute their surface integral contributions at the same time.
        When they both attempt to simultaneously add that contribution to $\mathbf{c}_{i,j}$, that memory becomes corrupted.
        We attempted to use the \lstinline|atomicAdd| operator, in order to bypass race conditions by serializing conflicting addition operations in this kernel.
        Atomic operators are known, however, to significantly degrade performance; in our implementation, runtime while using atomic operators increased by a factor of nine.
        In order to avoid using atomic operators, we chose instead to store each term separately in GPU global memory variables \lstinline|rhs_surface_left| and \lstinline|rhs_surface_right| and combine them later; see Section \ref{sec:rhs}.  
        The former variable stores the contribution in (\ref{eq:edgeint}) for the $\Omega_l$, the latter variable stores the contribution in (\ref{eq:edgeint}) for $\Omega_r$.
        The data organization of these arrays are the same as the $\mathbf{C}$ array described in Section \ref{sec:ds}.

        The parallel computation of  (\ref{eq:edgeint}) is displayed in pseudocode in Algorithm \ref{alg:eval_surface}.  We create one thread for each edge $e_i$, $i\in[1,N_s]$, tasked with computing (\ref{eq:edgeint}) over that edge.
        Each thread reads the coefficients of its left and right elements ${}^t\mathbf{c}_l[j,m]$ and ${}^t\mathbf{c}_r[j,m]$ for $j = [1.. N_p],~m=[1..M]$.
        
        The unstructured nature of our mesh precludes us from sorting the coefficients in memory to allow coalesced reads in both the volume integral kernel and the surface integral kernel.
        We may, however, sort the edge index list to enable coalesced reads of either ${}^t\mathbf{c}_{l}$ or ${}^t\mathbf{c}_{r}$ but not both.

        If edge $e_i$ lies on a computational domain boundary, a ghost state $\mathbf{U}_g$ is created and is assigned to $\mathbf{U}_r$.
        This ghost state depends on the type of boundary conditions used, e.g. solid, Dirichlet or far field boundaries.
        We sort edges so that all boundary edges appear first in our edge list.
        This avoids warp divergence, as the boundary edges with boundaries of the same type will be grouped in the same warp and treated similarly.  That is, all threads will take the same branch of the if-statement in Algorithm 2 (line 16).  This is true for all warps with the exception of a few; the precise number depends on the number of boundary condition types prescribed.  It is entirely possible to implement boundary specific surface integration kernels, but this would increase code complexity without noticeable gains to performance.

        Each thread must read $2 \times N_p \times M$ solution coefficients to compute $\mathbf{U}_l$ and $\mathbf{U}_r$.  Doing so requires a large amount of thread private memory, with registers quickly spilling into local memory.  For all orders of approximation $p$, \lstinline|eval_surface| uses 63 registers (the maximum number available), and an amount of local memory dependent on $p$; see Section \ref{sec:occupancy}, Table \ref{table:occupancy}.

        \begin{algorithm}[tbp] 
            \caption{\lstinline|eval_surface|}
            \label{alg:eval_surface}
            \begin{algorithmic}[1]
            \Require  \parbox[t]{\dimexpr\linewidth-\algorithmicindent}{A grid of $\lceil N_s/256 \rceil$ blocks with $256$ threads per block.  Each thread $i$ computes on one edge $e_i$.\strut}
            \Require \parbox[t]{\dimexpr\linewidth-\algorithmicindent}{Global Memory Inputs: thread $t_i$ receives edge data for $e_i$ (connectivity, normals, vertices), and the solution coefficients on $\Omega_l$ and $\Omega_r$ that share $e_i$, i.e. ${}^g\mathbf{C}[l,j,m]$ and ${}^g\mathbf{C}[r,j,m]$.\strut}
            \Require \parbox[t]{\dimexpr\linewidth-\algorithmicindent}{Constant Memory Inputs: thread $t_i$ receives ${}^c\mathbf{\Phi_q}$, and surface quadrature weights ${}^cw_k$.\strut}
                        \Ensure \parbox[t]{\dimexpr\linewidth-\algorithmicindent}{ Outputs: the surface integral contribution for edge $e_i$: ${}^g\text{\lstinline|rhs_surface_left|}[l,j,m]$, ${}^g\text{\lstinline|rhs_surface_right|}[r,j,m], j \in[1..N_p]$ and $m \in [1..M]$\strut}
            
			\State {$i \gets $ thread index}
            \State Load the left and right element number, $l$ and $r$ respectively, for edge $e_i$
            \State Load other relevant side information (normal, edge vertices, and edge length ${}^g\text{\lstinline|len|}$) into thread local memory
         \\   

            \State ${}^{g}\text{\lstinline|rhs_surface_left[l,j,m]|} \underset{ (j,m)}\gets 0$ 
            \State ${}^{g}\text{\lstinline|rhs_surface_right[r,j,m]|}\underset{ (j,m)}\gets 0$

            \If {$\Omega_r\neq$ boundary element}

            \State ${}^{t}\mathbf{c_l}[j,m]\underset{ (j,m)}\gets {}^{g}\mathbf{C}[l,j,m]$ 
            \State ${}^{t}\mathbf{c_r}[j,m]\underset{ (j,m)}\gets {}^{g}\mathbf{C}[r,j,m]$ 

            \Else

            \State ${}^{t}\mathbf{c_l}[j,m]\underset{ (j,m)}\gets {}^{g}\mathbf{C}[l,j,m]$

            \EndIf
\\
            \ForAll {surface integration points $k$ of edge $e_i$}
            \State ${}^{t}\mathbf{u_l}[m]\underset{ (m)}\gets {}^{t}\mathbf{c_l}[j,m]\underset{ (j)}\cdot {}^{c}\mathbf{\Phi_{q_l,k}}[j]$
            \If {$\Omega_r == $ boundary element}
			\State ${}^{t}\mathbf{u_r}[m]\underset{ (m)}\gets {}^d\text{\lstinline|eval_boundary|}(\mathbf{{}^{c}r_{q_r,p+1-k}})$ \Comment{Evaluate ghost state}
            \Else
            \State ${}^{t}\mathbf{u_r}[m]\underset{ (m)}\gets {}^{t}\mathbf{c_r}[j,m]\underset{ (j)}\cdot {}^{c}\mathbf{\Phi_{q_r,p+1-k}}[j]$
            \EndIf
            \State ${}^{t}\mathbf{F_n}[m] \underset{ (m)}\gets {}^d\text{\lstinline|riemann_solver|}({}^{t}\mathbf{u_l},{}^{t}\mathbf{u_r})$

			\\
			 \State \parbox[t]{\dimexpr\linewidth-\algorithmicindent}{$\text{\lstinline|rhs_surface_left[l,j,m]|} \underset{(j,m)} \gets \text{\lstinline|rhs_surface_left[l,j,m]|}-{}^t\text{\lstinline|len|} {}^{c}w_{q_l, k}{}^{t}\mathbf{F_n}[m]{}^{c}\mathbf{\Phi_{q_l,k}}[j]/2$ \strut}
			 
            \State \parbox[t]{\dimexpr\linewidth-\algorithmicindent}{$\text{\lstinline|rhs_surface_right[r,j,m]|} \underset{ (j,m)}\gets \text{\lstinline|rhs_surface_right[r,j,m]|}+$\\${}^t\text{\lstinline|len|}{}^{c}w_{q_r, k}{}^{t}\mathbf{F_n}[m]{}{}^{c}\mathbf{\Phi_{q_r,p+1-k}}[j]/2$ \strut}
            \EndFor
            \end{algorithmic}
            
        \end{algorithm} 

        \subsubsection{Right-Hand Side Evaluation Kernel (\lstinline|eval_rhs|)} \label{sec:rhs}

            The right-hand side evaluator kernel combines data from the three temporary storage variables ${}^g$\lstinline|rhs_surface_left|, ${}^g$\lstinline|rhs_surface_right|, and ${}^k$\lstinline|rhs_volume| to compute the right-hand side of equation (\ref{eq:2ddg0}).
            Each thread $t_i$ in the right-hand side evaluator kernel is assigned to one element $\Omega_i$.  The threads then combine the contributions from the surface and volume integrals for coefficients $\mathbf{c}_{i,j}$, $j = [1..N_p]$.
            The thread must determine if its element is seen as left or right for each of its edges.
            If the edge considers $\Omega_i$ a left element, the thread reads from ${}^g$\lstinline|rhs_surface_left|; on the other hand, if it considers $\Omega_i$ a right element, the thread reads from ${}^g$\lstinline|rhs_surface_right|.
            In both cases, the thread accesses for each $j = [1..N_p]$ and $m = [1..M]$ the appropriate memory locations of the three temporary storage variables and combines them to form the right-hand side of (\ref{eq:2ddg0}).

            As each thread must determine if $\Omega_i$ is considered a left or a right element for each of its edges, three boolean evaluations must be computed in this kernel.
            This introduces unavoidable warp divergence.
            As can be seen in Section \ref{sec:runtime}, Figure \ref{fig:gpuutilization}, this may contribute to a relatively large fraction of GPU time spent in \lstinline|eval_rhs| (compared to the amount of arithmetic), especially for lower orders of approximation.
            For all orders of approximation $p$, \lstinline|eval_rhs| uses 26 registers, and no local memory; see Section \ref{sec:occupancy}, Table \ref{table:occupancy}.
            
        \begin{algorithm}[H] 
            \caption{\lstinline|eval_rhs|}
            \label{alg:eval_rhs}
            \begin{algorithmic}[1]
            \Require A grid of $\lceil N/256 \rceil$ blocks, with $256$ threads per block
            \Require Global Memory Inputs: thread $t_i$ receives the volume contribution for element $\Omega_i$, surface contributions for its three sides, their edge connectivity data, and $|J_i|$.
            \Ensure Outputs: the right-hand side evaluation of (\ref{eq:2ddg0}) for element $\Omega_i$ into ${}^g\mathbf{C_{rhs}}$.

			\State {$i \gets $ thread index}

                 \State Load the determinant of the Jacobian ${}^g|J_i|$ and element edge $[e_{s_1},e_{s_2},e_{s_3}]$ data  
                 \\  

			\State ${}^g\mathbf{C_{\text{rhs}}}[i,j,m] \underset{(j,m)}\gets {}^g\mathbf{C_{\text{rhs}}}[i,j,m] +  {}^g\text{\lstinline|rhs_volume|}[i,j,m]/{}^t|J_i| $

            			\ForAll {$e_k$ of $\Omega_i$ where $k = [s_1,s_2,s_3]$ (the three edge indices of the sides of the element)}
            			
            			\If {$\Omega_i ==$ left element for $e_k$}

			\State \parbox[t]{\dimexpr\linewidth-\algorithmicindent}{ ${}^g\mathbf{C_{\text{rhs}}}[i,j,m] \underset{(j,m)}\gets {}^g\mathbf{C_{\text{rhs}}}[i,j,m] +  {}^g\text{\lstinline|rhs_surface_left|}[i,j,m]/{}^t|J_i| $ \strut}

			\Else
						\State \parbox[t]{\dimexpr\linewidth-\algorithmicindent}{ ${}^g\mathbf{C_{\text{rhs}}}[i,j,m] \underset{(j,m)}\gets {}^g\mathbf{C_{\text{rhs}}}[i,j,m] +  {}^g\text{\lstinline|rhs_surface_right|}[i,j,m]/{}^t|J_i| $\strut}
			\EndIf
			\EndFor

\end{algorithmic}
\end{algorithm}

       \subsubsection{Limiting Kernel} \label{sec:limiter}
        \begin{figure}[H]
            \caption{To limit the solution over $\Omega_i$, we evaluate the centroid values of surrounding elements $\Omega_a, \Omega_b$, and $\Omega_c$}
            \label{fig:limit}
            \centering
            \includegraphics{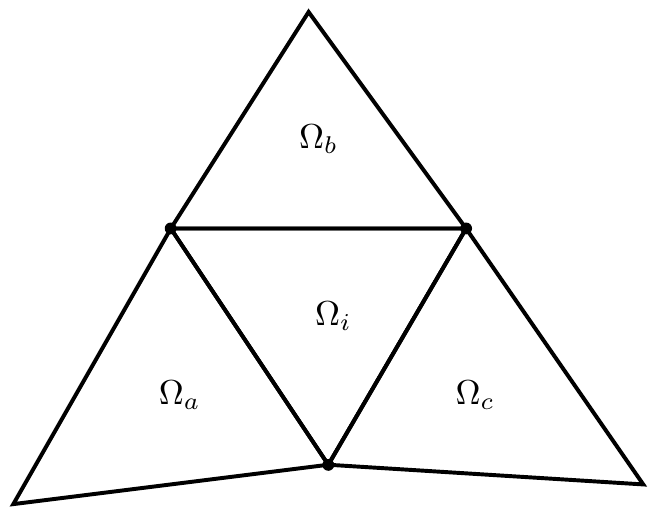}
        \end{figure}

        We implement the Barth-Jespersen limiter \cite{limiter} for linear $p = 1$ approximations.
        We aim to limit the maximum slope in the gradient of the scalar equation
        \begin{align}
            \label{eq:limit}
            U_i(\mathbf{r}) = \bar{U}_i + \alpha_i (\nabla U_i) \cdot (\mathbf{r} - \mathbf{r}_i),
        \end{align}
        by selecting a limiting coefficient $\alpha_i$. 
        In (\ref{eq:limit}), $\bar{U}_i$ is the average value of $U_i$ over $\Omega_i$ and $\mathbf{r}_i$ is the coordinate of the centroid of $\Omega_i$.
        Limiting systems of equations involves finding a separate $\alpha_i$ for each variable in the system.

        Suppose that element $\Omega_i$ is surrounded by elements $\Omega_a, \Omega_b$, and $\Omega_c$, as shown in Figure \ref{fig:limit}.
        We choose $\alpha_i$ so that $U_i$ introduces no new local extrema at the integration points on the boundaries relative to the averages on the three surrounding elements.
        We first evaluate $U_i$, $U_a$, $U_b$, and $U_c$ at their centroids.
        We then define the maximum centroid value
        \begin{align}
            \label{eq:limitmax}
            U_i^{\max} = \max\left\{U_i(\mathbf{r}_i), U_a(\mathbf{r}_a), U_b(\mathbf{r}_b), U_c(\mathbf{r}_c)\right\}
        \end{align}
        and minimum centroid value 
        \begin{align}
            \label{eq:limitmin}
            U_i^{\min} = \min\left\{U_i(\mathbf{r}_i), U_a(\mathbf{r}_a), U_b(\mathbf{r}_b), U_c(\mathbf{r}_c)\right\}.
        \end{align}

        Our implementation of this limiter operates element-wise.
        Each thread $t_i$ computes $\alpha_i$ to limit the slope of the approximation over a single element $\Omega_i$.
        Thread $t_i$ first computes $U_i^{\max}$ and $U_i^{\min}$ as in (\ref{eq:limitmax}) and (\ref{eq:limitmin}).
        Then, at each integration point $\mathbf{r}_{q,k}$ on the boundary of $\Omega_i$, thread $t_i$ computes $U_i(\mathbf{r}_{q,k})$ in order to compute $\alpha_{i,k}$.
        The smallest of the $\alpha_{i,k}$ values becomes the limiting constant $\alpha_i$.
        Finally, the coefficients $c_{i,2}$ and $c_{i,3}$ are multiplied by this $\alpha_i$.  
        This is repeated for each variable in the system.

        Each evaluation of $\alpha_i$ requires a significant number of boolean evaluations.
        As such, unavoidable warp divergence certainly inhibits performance. 
        The impact of this kernel will be shown numerically in Section \ref{sec:benchlim}.

\section{Computed Examples}

    \begin{table}[h]
    \centering
        \caption{GPU Specifications}
        \label{table:gpu}
        \begin{tabular}{|c|c|}
                       \hline
                   Device & NVIDIA GTX 580 \\ \hline
            		  Memory  & 3 GB GDDR5  \\
            CUDA Cores  & 512  \\
            Arithmetic Throughput (fp64)  & 197.6 GFLOP/s \\
            DRAM Memory Bandwidth  & 192.4  GB/s  \\
                        \hline
        \end{tabular}
    \end{table}

    We now present computed examples from this implementation of the DG method.
    Each example demonstrates solutions of Euler equations in two dimensions.
    Our simulations ran on the NVIDIA GTX 580, the specifications of which are detailed in Table \ref{table:gpu}.
    All tests were run on Ubuntu Linux 13.10 using CUDA 5.5, and code was compiled for devices of compute capability 2.0 using the \lstinline|-arch=sm_20| compiler flag.

    This implementation makes use of double precision floating point numbers whenever necessary.
    Our device's double precision arithmetic (fp64) throughput was calculated to be $1.544$ GHz/core $\times~512$ cores $\times~4/32$ FMAD\footnote{Fused Multiply Add} operations/clock cycle $\times ~2~$ FLOP\footnote{Floating Point Operation}/ FMAD operation $= 197.6$ FLOP/s.  Other architecture information was found in \cite{gtx580,cuda}.

    Mesh generation and solution visualizations were done using GMSH.
    All solutions were plotted using linear interpolation with no smoothing applied.
    The discontinuous nature of the numerical solution allows sharp jumps at isolines whenever solution values differ greatly between elements.

        The Euler equations describe the flow of an inviscid, isotropic, compressible fluid.
        In two dimensions, they are given by
        \begin{align}
            \label{eq:euler}
            \partial_t\begin{pmatrix}
                \rho \\
                \rho u \\
                \rho v \\
                E 
             \end{pmatrix}
             +
             \partial_x\begin{pmatrix}
                \rho u \\
                \rho u^2 + p \\
                \rho u v \\
                u(E + p) 
             \end{pmatrix}
             +
             \partial_y\begin{pmatrix}
                \rho v \\
                \rho u v \\
                \rho v^2 + p \\
                v(E + p)
             \end{pmatrix}
             =
             \mathbf{0},
         \end{align}
         where $\rho$ is the density, $u$ and $v$ are the velocity components, and $E$ is the energy.
         The variable $p$ in equation ($\ref{eq:euler}$) is the pressure given by an equation of state, which we choose to be
         \begin{align}
            p = (\gamma - 1) \left(E - \frac{\rho ||\mathbf{v}||^2_2}{2}\right),
         \end{align}
         for an adiabatic constant $\gamma$ and velocity vector $\mathbf{v} = (u, v)$.
         For air, we take $\gamma = 1.4$.

     \subsection{Supersonic Vortex}\label{sec:sv}


         \begin{figure}[tbp]
            \centering
            \caption{Example mesh and solution for the supersonic vortex test problem}
            \label{fig:advec2mesh}
           \begin{subfigure}[h!]{0.4\textwidth}
                \label{fig:cylmeshsmall}
                \includegraphics[width=\textwidth]{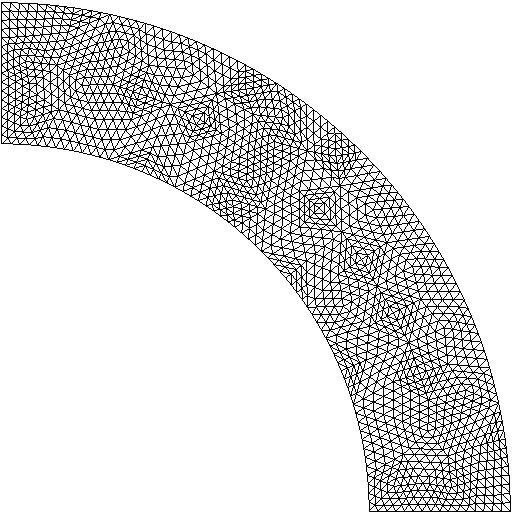}
                \caption{Mesh $C$}
            \end{subfigure}
            \quad
            \begin{subfigure}[h!]{0.4\textwidth}
                \label{fig:solC}
                \includegraphics[width=\textwidth]{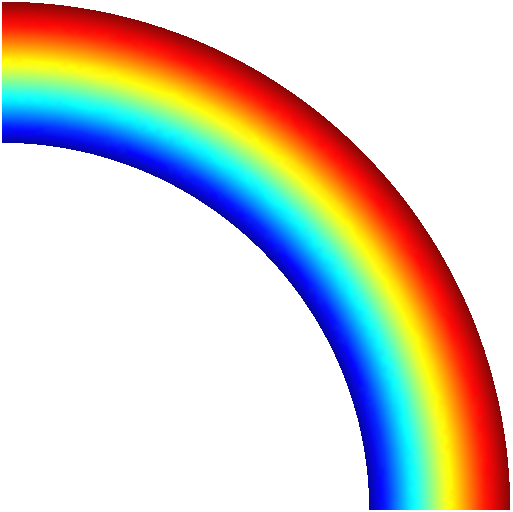}
                \caption{Solution on mesh $C$ for $p = 1$}
            \end{subfigure}
          \end{figure}

        \begin{table}[tb]
            \centering
            \caption{$L^2$ error in density and convergence rate $r$ for levels of $h$- and $p$-refinement for the supersonic vortex test problem}
            \label{table:svhp}
            \begin{tabular}{|l|llllllll|}
                \hline
                & \multicolumn{2}{|c|}{$p = 1$} & \multicolumn{2}{|c|}{$p = 2$} & \multicolumn{2}{|c|}{$p = 3$} & \multicolumn{2}{|c|}{$p = 4$} \\
                \hline
                Mesh& Error & $r$                   & Error & $r$                   & Error & $r$                   & Error & $r$ \\
                \hline
                $A$ & $4.934$E$-3$ & - &
                      $3.708$E$-4$ & - &
                      $8.695$E$-6$ & - &
                      $4.719$E$-7$ & - \\
                $B$ & $1.226$E$-3$ & 2.009 &
                      $6.003$E$-5$ & 2.627 &
                      $5.598$E$-7$ & 3.957 &
                      $1.887$E$-8$ & 4.644 \\
                $C$ & $3.267$E$-4$ & 1.908 &
                      $8.077$E$-6$ & 2.894 &
                      $3.237$E$-8$ & 4.645 &
                      $6.925$E$-10$ & 4.766 \\
                $D$ & $8.695$E$-5$ & 1.910 &
                      $1.043$E$-6$ & 2.953 &
                      $1.904$E$-9$ & 4.086 &
                      $2.189$E$-11$ & 4.983 \\
                \hline
            \end{tabular}
        \end{table}

         The supersonic vortex test problem models supersonic fluid flow through a curved quarter-cylinder tube.
         This problem has a known smooth analytical solution, which we prescribe as the initial conditions.
         We use curved reflecting boundary conditions, detailed in \cite{boundary}, along the curved edges and inflow and outflow boundary conditions at the inlet and outlet.
         We run the simulation until numerical convergence occurs, defined by
         \begin{align}
            \max_{i,j} \left\{|\mathbf{c}_{i,j}^{n+1} - \mathbf{c}_{i,j}^n|\right\} \leq 10^{-14}.
         \end{align}
 
         We use a convergence analysis to verify this implementation for nonlinear problems.
         We call our meshes $A$ through $D$ with mesh $A$ containing 180 elements.
         Meshes $B$ through $D$ were created by successive refinement of the previous mesh by splitting each triangle into four triangles, quadrupling the total number of elements with each refinement; mesh characteristics are presented in Table \ref{tab:benchmesh}.

         The $L^2$ error between the numerical steady state and analytical steady state are compared for each combination of $h-$ and $p-$refinement in Table \ref{table:svhp}.
         These convergence rates match theoretical convergence rates, verifying the accuracy of our implementation for nonlinear problems.
         Our results look both quantitatively and qualitatively similar to those found elsewhere; e.g., \cite{boundary}.

      \subsection{Double Mach Reflection}
         \begin{figure}[tbp]
             \caption{Computational domain $\Omega$ for the double Mach reflection test problem. The striped triangle represents the reflecting wedge. The shaded region on the left is the shock region $\mathbf{U}_s$ while the region on the right is the pre-shock condition $\mathbf{U}_q$.}
             \label{fig:dmdiagram}
             \centering
             \includegraphics[width=0.6\textwidth]{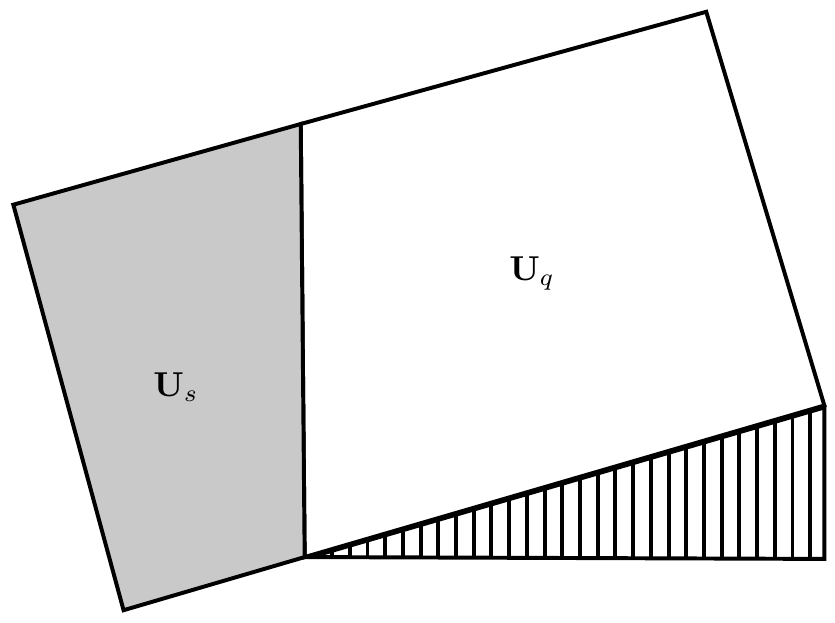}
         \end{figure}

        \begin{figure}[tbp]
            \centering
            \caption{Density for the double Mach reflection problem using $p = 1$}
            \label{fig:dmsol}
            \begin{subfigure}[h!]{0.95\textwidth}
                \centering
                \includegraphics[width=\textwidth]{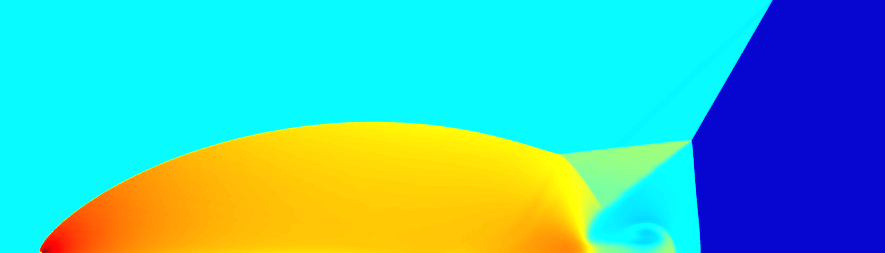}
                \caption{Density for mesh $C$}
            \end{subfigure}
            \quad
            \begin{subfigure}[h!]{0.95\textwidth}
                \centering
                \includegraphics[width=\textwidth]{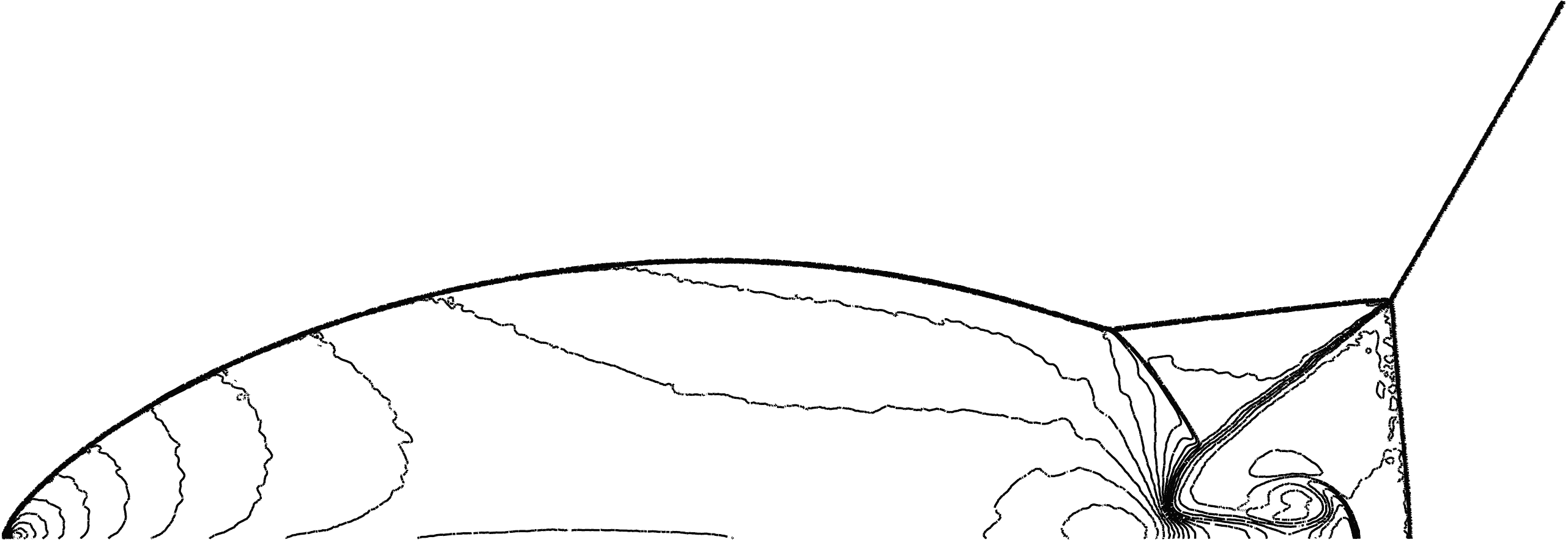}
                \caption{Density isolines for mesh $C$}
            \end{subfigure}
        \end{figure}
 
         The double Mach reflection test problem models a planar shock wave over a reflecting angled wedge.
         This is equivalent to modeling an angled shock wave moving over a straight reflecting wall.
         The reflections of the initial shock wave create additional shock waves and contact discontinuities during this simulation.
         
         We begin by sending a right-moving incident Mach 10 shock wave with propagation angle $\theta = 60^\circ$ from the $x$-axis; i.e., we assume the wedge has a half-angle of $30^\circ$.
         The computational domain $\Omega$ is defined by $x = [0,4], y = [0,1]$.
         The lower boundary in our domain models a reflecting wedge beginning at $x_0 = \frac{1}{6}$.

         We assume that the unperturbed flow's density and pressure are equal to 1.4 and 1, respectively.
         The after shock values are calculated to satisfy the Rankine-Hugoniot condition.  The left boundary condition sets an inflow with $\mathbf{U}_s$ as the parameter values.
        The boundary condition along the top of the domain keeps up with the speed of the incident shock wave to simulate the effects of an infinitely long wave.
        Along the top boundary, at integration points to the left of the shock wave, the exact values from $\mathbf{U}_s$ are prescribed, while points to right of the shock wave use the values from $\mathbf{U}_q$.
        The lower boundary condition prescribes the values of the shock $\mathbf{U}_s$ at $x \leq x_0$ and uses reflecting boundary conditions beyond to simulate a perfectly reflecting wedge.

       \begin{table}
            \caption{Performance of the double Mach reflection test problem}
            \label{table:dmruntime}
            \centering
            \begin{tabular}{|lcccc|}
            \hline
            Mesh &  Elements  & Memory & Runtime (min) & Timesteps\\
            \hline
            $A$ & 68,622  & 43.64 MB  &  0.71 & $3,961$\\
            $B$ & 236,964 & 176.48 MB & 8.71 & $12,120$\\
            $C$ & 964,338 & 717.82 MB & 68.32 &$23,160$ \\ 
            \hline
            \end{tabular}
        \end{table}
        Our test set runs over three unstructured triangular meshes of varying mesh sizes reported in Table \ref{table:dmruntime}.
        We compute the solution until $t = 0.2$ when the shock has moved nearly across the entire domain.
        Our solution is computed using $p = 1$ (linear) polynomials with the slopes limited using the Barth-Jespersen limiter.
        Mesh refinement is done by setting a smaller maximum edge length and creating a new mesh with GMSH.

        The density and density isolines at $t = 0.2$ for the most refined mesh $C$ are plotted in Figure \ref{fig:dmsol}.
        Our jet stream travels and expands as in similar simulations in \cite{doublemach}.
        The exact boundary condition at the top edge of our domain introduces small numerical artifacts which can be seen at the top of the domain behind the shock.

        Table \ref{table:dmruntime} also reports total runtime and memory costs for these simulations using the classical second-order Runge-Kutta time integration scheme.
        The simulation time, even for the very large meshes, is not prohibitive.
        Meshes of $C$'s size are typically too large to be run in serial implementations, usually requiring supercomputing time.
        In contrast, our GTX 580 completed this simulation in just over an hour.
        \section{Kernel Performance Characteristics}
In this section, the performance of our implementation is analyzed using tools available in the CUDA toolkit, such as the profiler \lstinline|nvprof|.  
For all the following analyses the tests were run on the supersonic vortex problem using the classical fourth order Runge-Kutta time integrator.  Unless otherwise stated, the solution was not limited and mesh E was used to ensure device saturation.  
We report the relative timing of the three main compute kernels, occupancy considerations, and memory bandwidth and arithmetic throughput metrics.

\subsection{Kernel Runtime Distribution} \label{sec:runtime}
        \begin{figure}
            \centering
                         \caption{Relative GPU utilization of major and minor kernels during an RK4 timestep}
             \includegraphics[width=1\textwidth]{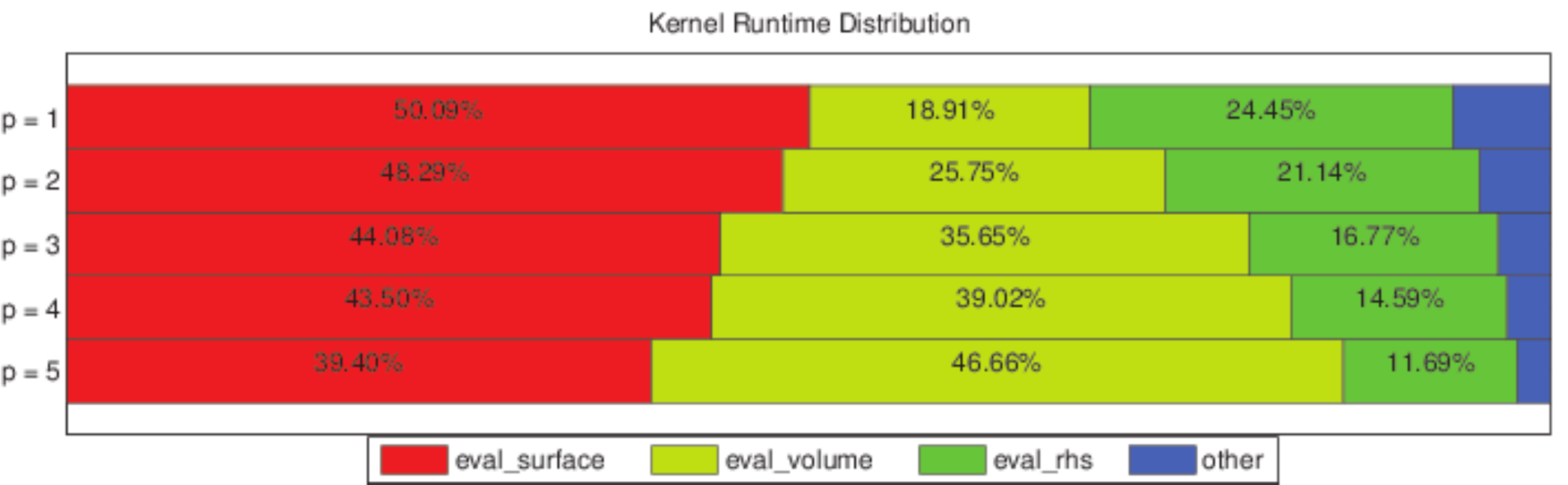}
             \label{fig:gpuutilization}
         \end{figure}
 \label{sec:gpuutilization}

We present now the relative time spent by the GPU in the surface, volume, and right-hand side evaluation kernels during a typical timestep.  Measurements were completed by instrumenting the algorithm with calls to \lstinline|cudaEventElapsedTime|.  GPU runtime was averaged over 500 timesteps.  The results are presented in Figure \ref{fig:gpuutilization}.  Minor kernels determining stable timesteps, completing scalar-vector multiplications, vector-vector sums, and device-to-host memory transfers are grouped together in the `other' category.

It is observed that for low orders of approximation, the surface integration kernel is the most time consuming.  For higher order approximations, the volume integration kernel takes over.  This is because the number of volume integration points grows faster than the number of surface integration points. Compare 3 volume integration points to 6 surface integration points for $p = 1$ on one triangle, and 16 volume integration points versus 15 surface integration points for $p = 4$.

 \subsection{Occupancy} \label{sec:occupancy}
The occupancy metric provides an indication of how well the kernel will be able to hide memory latencies.   It is defined as the ratio of active warps resident on a multiprocessor to the maximum possible number of active warps the multiprocessor can handle \cite{bestpractices}.

  The occupancy achieved for each kernel varied very little with the order of approximation $p$.  This is because the compiler placed large arrays whose size depends on $p$ directly into local memory.  For this reason, register usage and occupancy were not affected by the order of approximation.  This has been verified, but not reported for brevity.

       \begin{table} [h!]
            
            \centering
            \begin{tabular}{|c|c|c|c|c|}
            \hline
            Kernel & Achieved Occupancy & Theoretical Occupancy & Register Usage \\
            \hline
            \lstinline|eval_volume| & 0.321687 & 0.33 & 43  \\
            \hline
            \lstinline|eval_surface| & 0.322379 & 0.33 & 63 \\
            \hline
            \lstinline|eval_rhs|& 0.625838 &0.67 & 26  \\
            \hline
            \end{tabular}
            \caption{Kernel cccupancy and register usage for $p=1$  and $256$ threads per block}\label{table:occupancy}
        \end{table}
In Table \ref{table:occupancy}, occupancy and register usage are presented for $p=1$.  
It can be seen that both \lstinline|eval_volume| and \lstinline|eval_surface| use a large number of registers, reducing the possible theoretical occupancy.  
Setting a global register maximum using the \lstinline|maxrregcount| compiler flag would allow for greater occupancy.  This has been implemented only to find that kernel execution speed was hindered.  
This is likely because the increased occupancy does not manage to hide the latencies incurred by the increased memory transactions to local memory in the caches and possibly even DRAM.
         
\subsection{Memory Bandwidth}
Most GPU kernels are bound by the available memory bandwidth; ensuring efficient loads and stores from global memory is therefore critical to overall performance.  We now examine the DRAM memory bandwidth of the main compute kernels, as defined by the following relation

\begin{equation*}
\text{BW} = \text{\lstinline|dram_read_throughput|} + \text{\lstinline|dram_write_throughput|}.
\end{equation*}

\noindent
In Figure \ref{fig:mem}, it is observed that the DRAM throughput is relatively constant remaining between $70$ and $80\%$ of the theoretical maximum for all three kernels, indicating that our implementation is bandwidth-bound.  Comparing with the linear nodal DG-GPU implementation in \cite{dggpu1,dggpu2} we similarly see a high memory bandwidth utilization.

With the kernels using the majority of the hardware bandwidth available, we now consider the efficiency of these transfers.  In CUDA, when a warp of 32 threads completes a coalesced read of 32 double precision floating point numbers, this results in two 128B cache line requests.  This is the most efficient access possible, as all the loaded data is used in the kernel.  If a suboptimal memory access pattern is used, the bandwidth will become polluted with unused data.  
This is because more than 2 cache lines will be requested, and some unnecessary data will be moved.  

The metric of interest is therefore the ratio between the number of read/write transactions per request.  It is given by \lstinline|nvprof|'s \lstinline|gld_transactions_per_request| and \lstinline|gst_transactions_per_request|, respectively.  
Both quantities should be as close to 2 as possible for double precision floating point transfers \cite{mich}.  
For all kernels, the write transactions per request, $\text{\lstinline|gst_transactions_per_request|}$, is about $2$; for the surface integration and right-hand-side evaluation kernels, the read transactions per request, $\text{\lstinline|gld_transactions_per_request|}$, is between $\sim 2$ and $\sim 3$, due to irregular read patterns.  For the rest of the kernels, this metric is about 2.

\subsection{Arithmetic Throughput}
\begin{figure}[tbp]
            \centering
            \caption{Kernel arithmetic throughput and memory bandwidth.   Net throughput and bandwidth were calculated using a time-weighted average over all major and minor kernels}
            \label{fig:metrics}
            \includegraphics[width=0.55\textwidth]{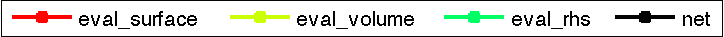}
           \begin{subfigure}[h!]{0.45\textwidth}
            \centering
             \includegraphics[width=1\textwidth]{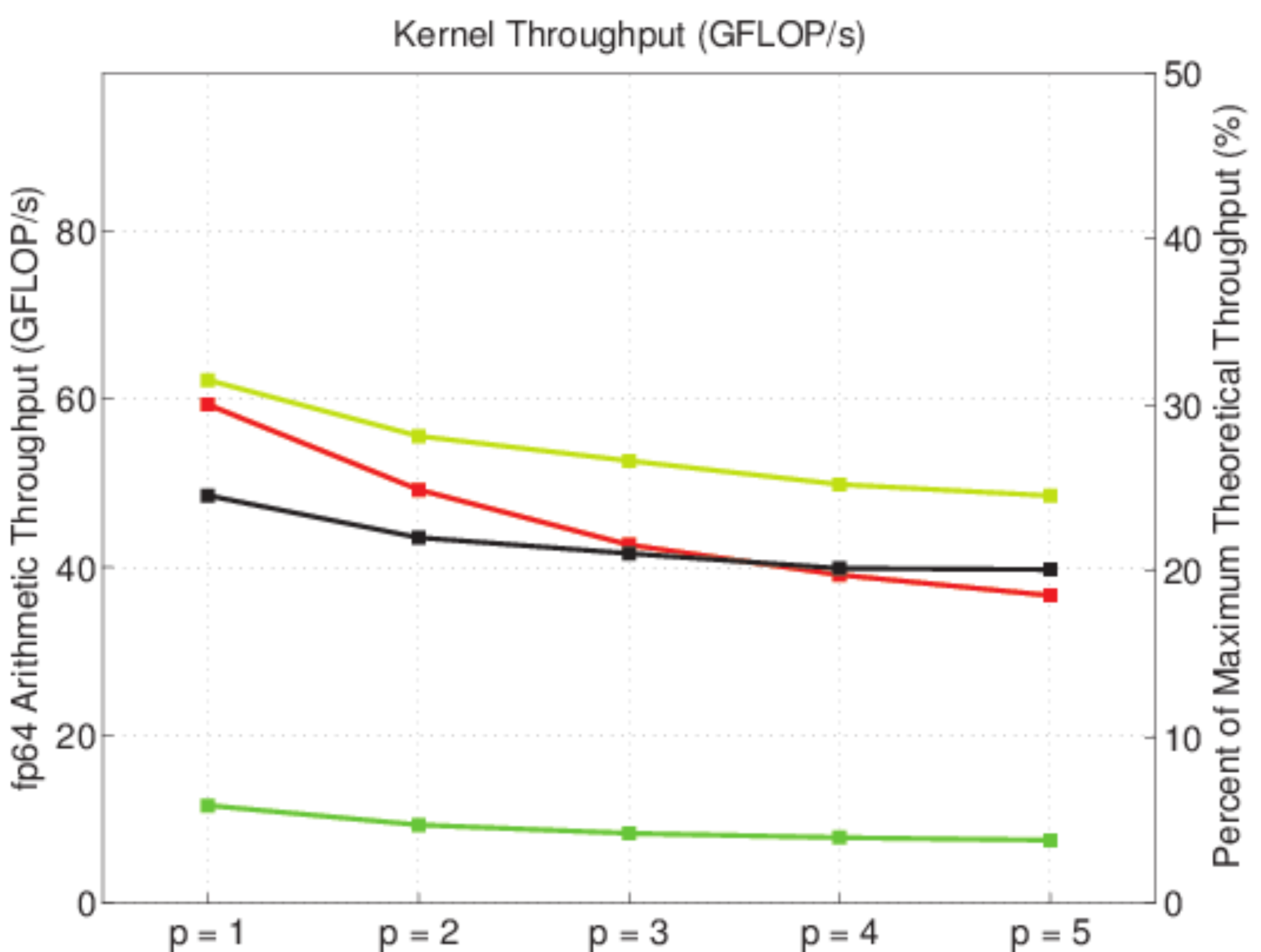}
             \caption{Kernel Arithmetic Throughput (GFLOP/s)}
             \label{fig:flop}
            \end{subfigure}
            \quad
            \begin{subfigure}[h!]{0.45\textwidth}
            \centering
             \includegraphics[width=1\textwidth]{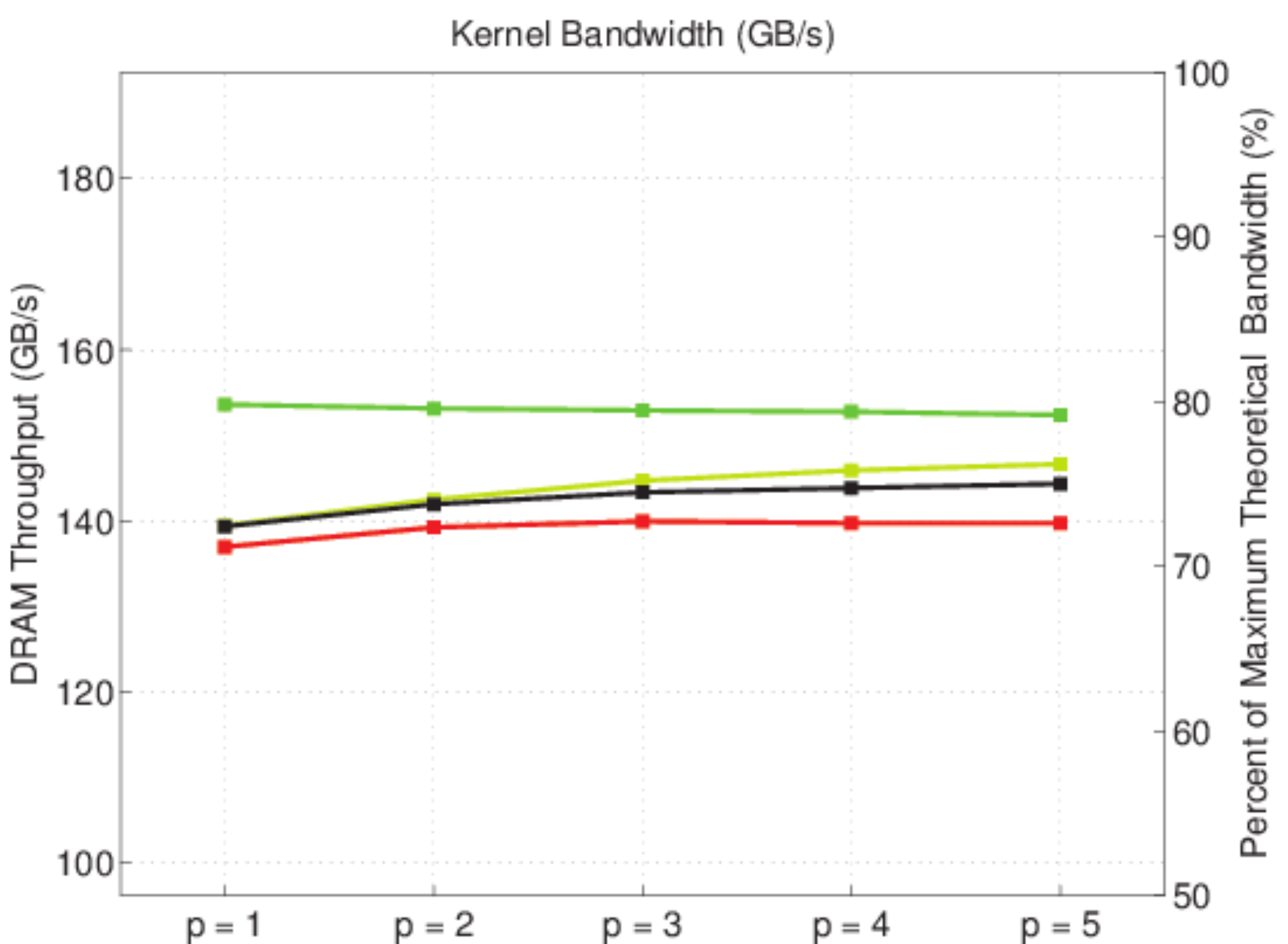}
             \caption{Kernel Memory Bandwidth (GB/s)}
             \label{fig:mem}
            \end{subfigure}
          \end{figure}

Next, we assess the arithmetic throughput of the three main kernels.  This metric is defined as
\begin{equation*}
\text{FLOP/s} = \frac{\text{\lstinline|flops_dp|}}{t},
\end{equation*}
where \lstinline|flops_dp| is the number of double precision FLOPs measured by \lstinline|nvprof| and $t$ is the average kernel runtime described in Section \ref{sec:gpuutilization}.  Measured FLOP throughput is reported in Figure \ref{fig:flop}.

For the main compute kernels \lstinline|eval_volume|, \lstinline|eval_surface|, and \lstinline|eval_rhs|, it is observed that the arithmetic throughput somewhat decreases with increasing $p$. 
 This is unexpected, as one would assume the higher arithmetic density of the kernels would be able hide memory access latencies.
 Being unable to hide latencies could be symptomatic of many things, i.e. low occupancy or a large number of dependent instructions.
 If we choose a number of threads per block yielding higher occupancy than those cited in Table \ref{table:occupancy}, the execution speed of the kernels does not improve according to performance tests.
  Thus, we can eliminate this possibility.
Each thread in the volume and surface kernels completes a large amount of operations, many of which are dependent on one another, i.e. evaluating the surface and volume integrals is contingent on a flux evaluation, which in turn depends on both the loading of the solution coefficients and the evaluation of $\mathbf{U}$ at the integration points.  
Too many instruction dependencies would have a detrimental effect on instruction level parallelism and, consequently, the arithmetic throughput of the kernel. 
 A third possible cause is that we have an increase in the amount of data accesses per thread, without an equivalent increase in memory bandwidth.  Threads are therefore requesting a greater amount of data, without a change in the rate at which it is delivered.

Unlike our implementation, Kl\"ockner et al. \cite{dggpu1,dggpu2} report an arithmetic throughput which generally increases with the order of approximation.  
	For $p=1$, we achieve a net arithmetic throughput of 25\% of our device's maximum.  This is much greater than the net arithmetic efficiency achieved by Kl\"ockner's implementation of approximately 5\%.  Further, when $p = 5$, our net throughput decreases to 20\% of our device's maximum.  This is just below the net arithmetic efficiency achieved by Kl\"ockner et al. of approximately 21\%.
	We did not complete simulations for $p$ greater than 5, but if we continue the trend, our implementation will exhibit lower arithmetic efficiency than Kl\"ockner et al.

The throughput of the \lstinline|eval_rhs| kernel is likely low as it has low arithmetic intensity, exhibits uncoalesced memory loads, and experiences some warp divergence, see also Section \ref{sec:rhs}.


    \section{Benchmarks}

        We now present the benchmarks for this implementation.
        First, we present how application performance scales with mesh sizes.  Next, we examine the performance degradation introduced by the use of a limiter.

            \begin{table}[H]
                \centering
                \caption{Mesh sizes for the supersonic vortex test problem used for benchmarking}
                \label{tab:benchmesh}
                \begin{tabular}{|l|cccccc|}
                    \hline
                    Mesh     & $A$ & $B$ &   $C$ &   $D$ &    $E$ &    $F$ \\
                    \hline
                    Elements & 180 & 720  & 2,880 & 11,520 & 46,080 & 184,320\\
                    Edges    & 293 & 1126 & 4,412 & 17,464 & 69,488 & 277,216\\
                    \hline
                \end{tabular}
            \end{table}

       \subsection{Scaling}

                  \begin{figure}[tbp]
            \centering
            \caption{GPU execution time and FLOP counts for meshes A to F and orders of approximation $p = 1$ and $p = 4$}
           \begin{subfigure}[h!]{0.45\textwidth}
            \centering
             \includegraphics[width=1\textwidth]{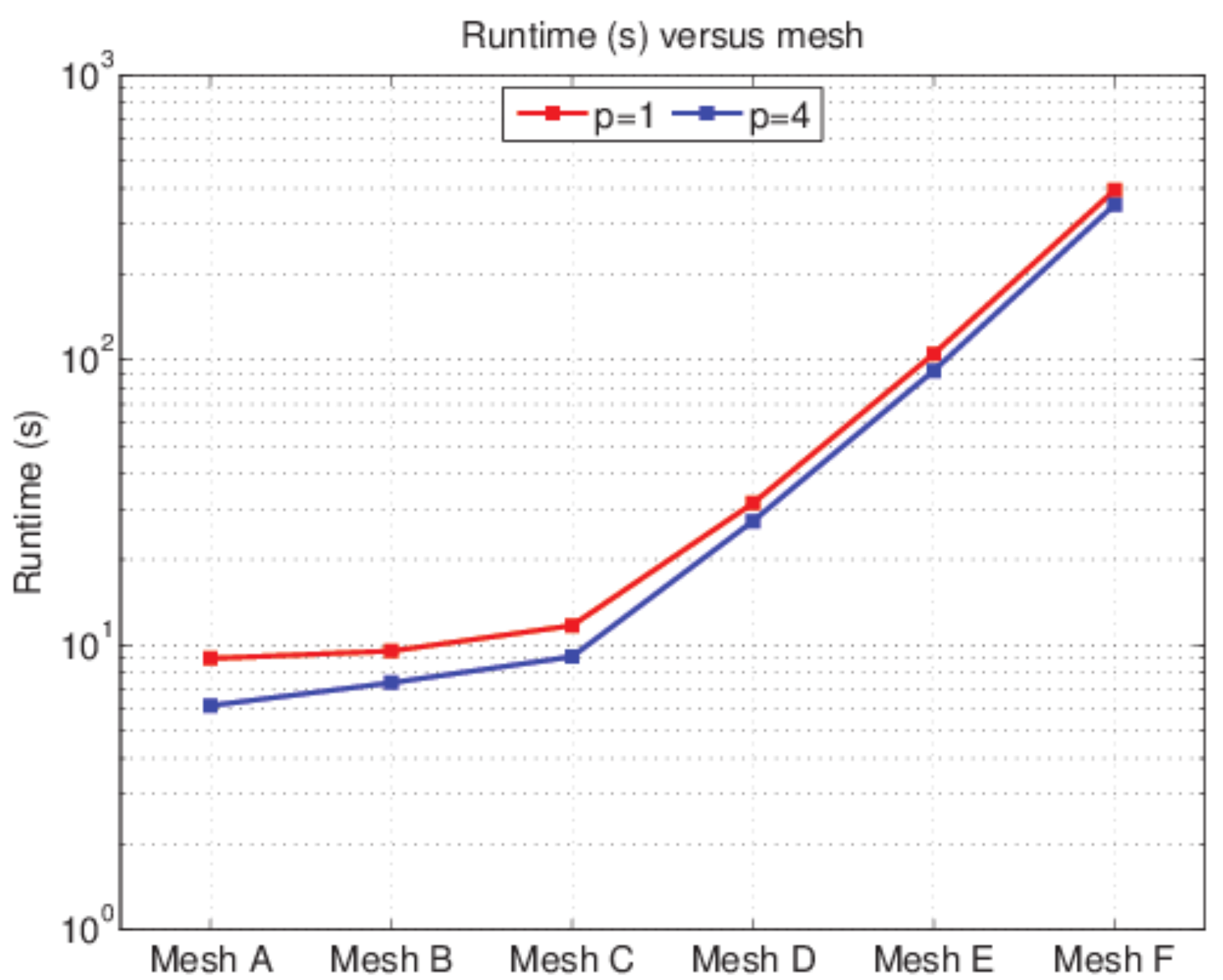}
             \caption{GPU execution time for $p = 1$ for 10,000 timesteps and $p = 4$ for 1,000 timesteps}
             \label{fig:scaling}
            \end{subfigure}
            \quad
            \begin{subfigure}[h!]{0.45\textwidth}
            \centering
             \includegraphics[width=1\textwidth]{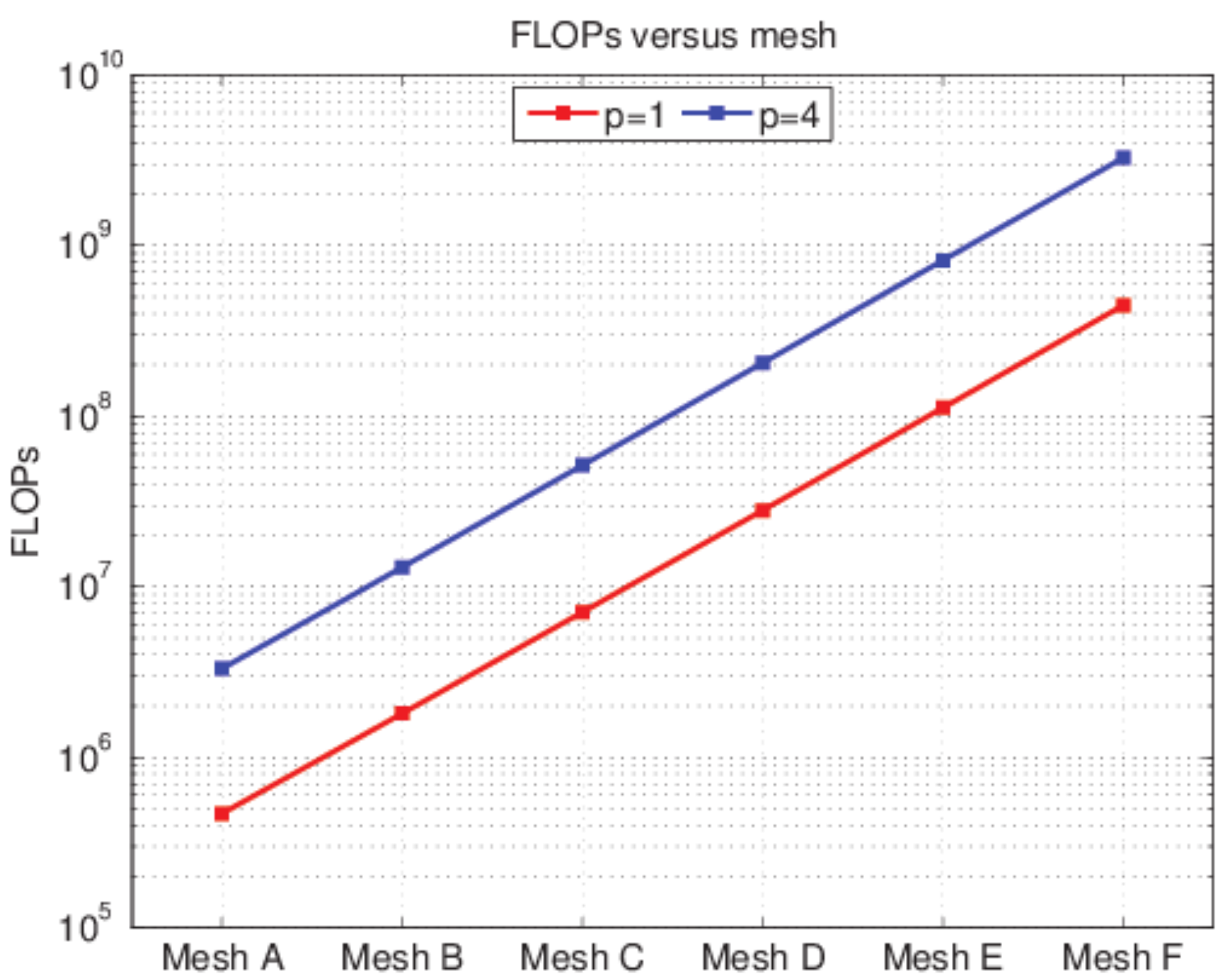}
             \caption{Total number of FLOPs executed during a single RK stage for each mesh by the kernels \lstinline|eval_surface|, \lstinline|eval_volume| and \lstinline|eval_rhs|}
             \label{fig:execflops}
            \end{subfigure}
          \end{figure}
            We first demonstrate the scalability of our implementation by measuring performance at device saturation.
            
            Device saturation requires a sufficiently large number of threads to be running simultaneously.
            As the total number of simultaneously running threads depends on the size of the mesh, we reach device saturation by computing over larger meshes.
    
            We first fix $p = 1$ and compute the supersonic vortex test problem described in Section \ref{sec:sv} over 10,000 timesteps on meshes $A$ through $F$ from Table \ref{tab:benchmesh}.
            Next, we fix $p = 4$ and repeat the test using 1,000 timesteps over the same meshes.
            The computation run times for these two tests are displayed in Figure \ref{fig:scaling} along with corresponding FLOP counts in Figure \ref{fig:execflops}.

            We see roughly an order of magnitude difference in execution times between $p = 1$ and $p = 4$, corresponding to approximately an order of magnitude increase in the number of FLOPs executed.
            This demonstrates that our implementation does indeed scale well, as an increase in problem size and complexity does not reduce execution efficiency.
            We also note that until we reach mesh $C$, execution time does not increase linearly as not enough threads are created to saturate the device.  Because the surface and volume contributions can be computed simultaneously, running \lstinline|eval_surface| and \lstinline|eval_volume| concurrently with streams would permit device saturation on smaller meshes.  For larger meshes however, this optimization would be of no benefit.

        \subsection{Limiting}\label{sec:benchlim}
        
            \begin{figure}[tpb]
               \caption{GPU execution time with $p = 1$ and 10,000 timesteps with and without limiting}
               \label{fig:limspeedup}
                \centering
               \begin{subfigure}[h!]{.8\textwidth}
                    \centering
                    \includegraphics[width=0.7\textwidth]{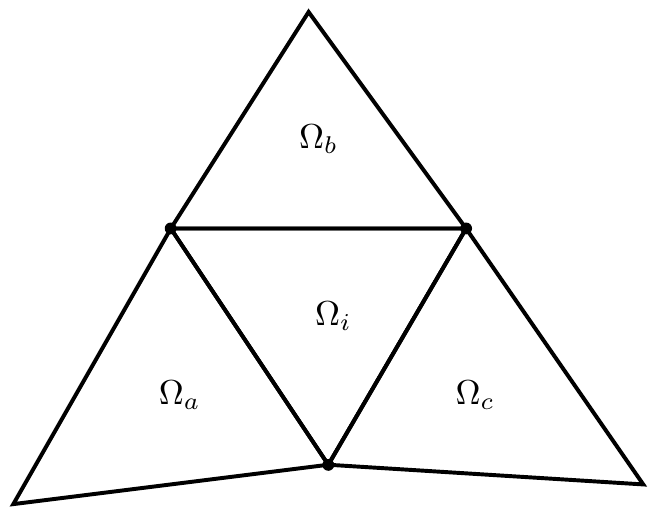}
                \end{subfigure}

            \end{figure}

            We now benchmark our implementation's performance with the Barth-Jespersen limiter described in Section \ref{sec:limiter}.
            
            Using the same supersonic vortex test problem with $p = 1$, we compare execution time with and without limiting over 10,000 timesteps.  We observe that though there is a degradation in performance, it is not prohibitory; on the mesh sizes examined, the limiter only slows execution time by a maximum $\sim 15\%$.

\section{Conclusion}
    
    We have presented a DG-GPU solver that achieves comparable device utilization to \cite{dggpu1,dggpu2} without the need for variable padding, shared memory or empirical testing to determine work partitioning.
    For low orders of approximation on two-dimensional nonlinear problems, our implementation exhibits comparable memory bandwidth and better arithmetic efficiency than  \cite{dggpu1,dggpu2}.  As the order of approximation reaches $p=5$, both implementations have comparable performance.
    
    Our solver easily computes the double Mach reflection test problem with nearly one million elements on the GTX 580 in about an hour.
    The three gigabytes of video memory on this device can compute linear approximations on meshes of around four million triangles.
    New GPU hardware containing even more video memory can allow us to tackle even larger problems.

    In future developments, we aim to reduce our memory usage and increase parallelism.
    Currently, we require three extra temporary storage variables to compute the right-hand side in (\ref{eq:2ddg0}).
    By eliminating the race conditions through edge list partitioning, we could add each contribution to the right-hand side as we compute them.
    This would involve developing an edge coloring algorithm, which is not straightforward for complicated meshes, e.g. a mesh exhibiting gaps such as a mesh around an airfoil.

    Further, adaptive mesh refinement would be a natural extension and fit into our implementation in a straightforward way.
    It would be very useful to extend our algorithm for use on higher order geometric elements and analyze the incurred impact on performance.  
    The main issues on curved mesh elements are the nonlinearity of mappings and the need for higher order integration rules.  
    We would also need to invert mass matrices which would no longer be orthogonal.  
    One efficient recently proposed approach to this last issue would be to introduce modified basis functions that preserve orthogonality on curved elements \cite{warburton2013}.  
    Though curved elements require more computational effort, the mesh can be sorted so they can be treated simultaneously as we currently do for boundary edges.  High order elements are usually used only along geometric boundaries, thus they comprise a small portion of the mesh and should not affect performance significantly.

    Finally, we intend to extend support to three-dimensional problems.

\section{Acknowledgment}
This research was supported in part by Natural Sciences and Engineering Research Council (NSERC) of Canada grant 341373-07.
   
\section*{Bibliography}
{\def\section*#1{} 
\bibliographystyle{unsrt}
\bibliography{paper}{}
}
\end{document}